\documentclass[preprint,3p,times,twocolumn]{elsarticle}

\usepackage{amssymb}
\usepackage{amsmath}
\usepackage{mathtools,cuted}
\usepackage[english]{babel}
\usepackage{hyperref}
\usepackage{xcolor}
\usepackage[normalem]{ulem}
\usepackage{algpseudocode} 

\begin{document}
\begin{frontmatter}
\title{SOQCS: A Stochastic Optical Quantum Circuit Simulator}

\author[a,b]{Javier Osca\corref{corr}\fnref{email1}}
\cortext[corr]{Corresponding author.}
\fntext[email1]{javier.oscacotarelo@mu.ie}
\author[a]{Jiri Vala\fnref{email2}}
\fntext[email2]{jiri.vala@mu.ie}
\address[a]{Department of Theoretical Physics, Maynooth University, Ireland}
\address[b]{Tyndall National Institute, University College Cork, “Lee Maltings,” Dyke Parade, Cork, Ireland}

\begin{abstract}
We present Stochastic Optical Quantum Circuit Simulator (SOQCS) C++/Python library for the simulation of quantum optical circuits, and we provide its implementation details. SOQCS offers a framework to define, simulate and study quantum linear optical circuits in the presence of various imperfections. These come from partial distinguishability of photons, lossy propagation media, unbalanced beamsplitters and non-ideal emitters and detectors for example. SOQCS is developed as a series of different modules which provide quantum circuits, different simulator cores and tools to analyze the output. Quantum circuits can be defined from basic components, including emitters, linear optical elements, delays and detectors. Post-selection can be configured straightforwardly as part of detector definitions. An important attribute of SOQCS is its modularity which allows for its further development in the future.
\end{abstract}

\end{frontmatter}

\section{Introduction}
\label{Intro}
\begin{figure}[h]
\includegraphics[width=0.5\textwidth]{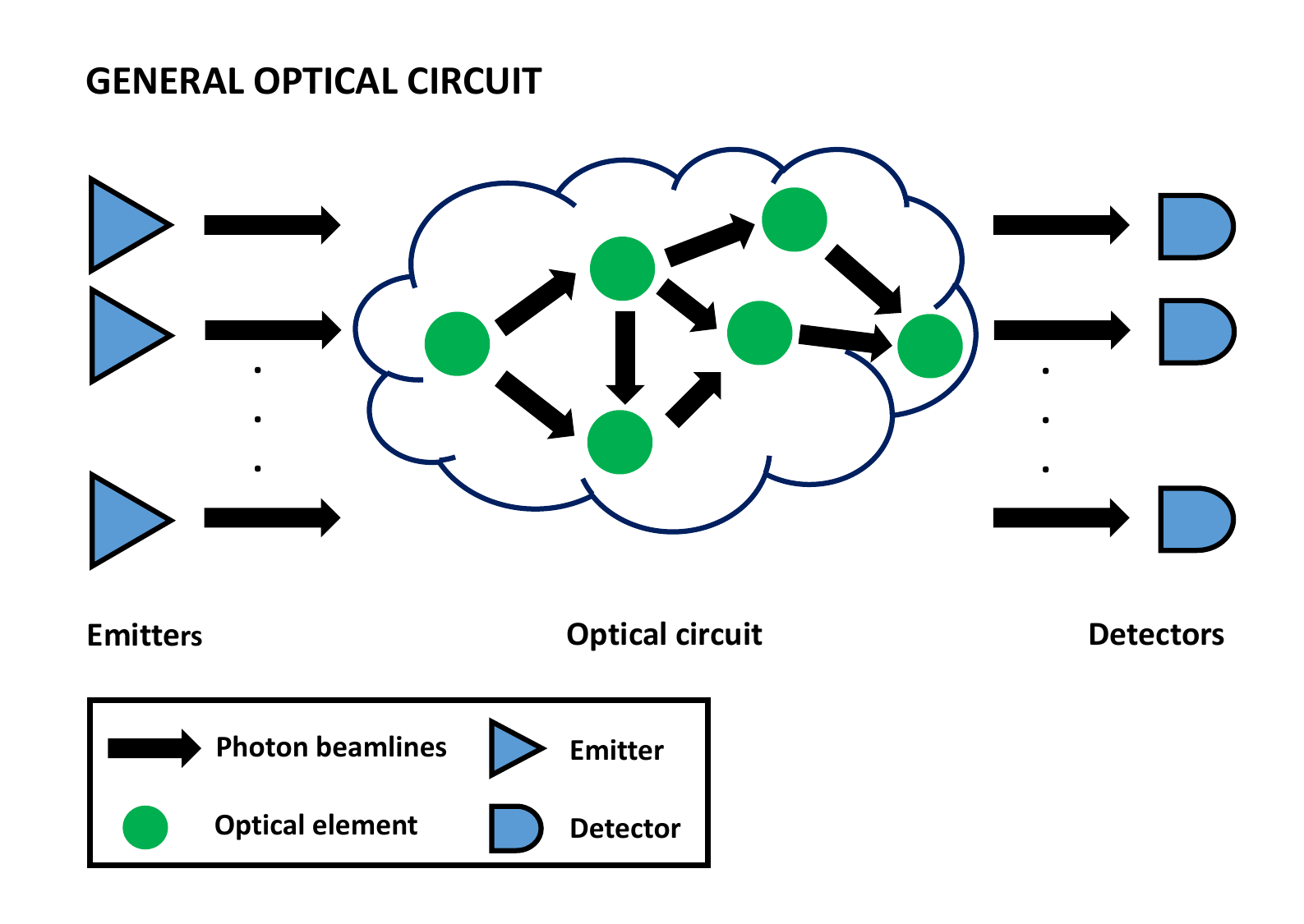}
\caption{Schematic of a quantum optical circuit. In the context of this article an optical circuit is defined as a set of interconnected linear optical elements, complemented by additional
operations such as measurement and delays.}
\label{F2}
\end{figure}

Quantum computing promises a considerable speed up of certain computational problems \cite{QComp,QComp2}. It has been shown that universal quantum computation can be attained using  optical circuits with linear elements and post-selection \cite{Linear1,Linear2,KLM}. A state of a quantum bit or qubit can be easily encoded using photonic quantum states. The initial photonic state is created by means of emitters and transformed by the optical circuit into the final state that is measured using detectors. A quantum optical circuit capable of complex operations is defined in this context as a network of simpler linear optical elements, such as phase shifters and beamsplitters. The schematic of a general optical circuit is illustrated in fig. \ref{F2}. 

There has been a considerable development of software platforms for quantum computing in the last few years. For example, IBM introduced the open source environment Qiskit \cite{Qiskit}. Also, tools such as Openquasm \cite{QASM}, Cirq \cite{Circ}, Quil \cite{Quil} and QP \cite{QP} concern the simulation of the logical operations of quantum computers. Additionally, there are platforms like D-Wave development kit \cite{Dwave} that are more oriented to interface with specific hardware. On the other hand, software focused on simulation and interface with optical platforms is primarily centered on continuous variable models \cite{Strawberry,FPAQS}.

In this paper we present the implementation details of the C++ and Python library SOQCS \cite{SOQCS} for the simulation of quantum optical circuits where photonic states are represented using a Fock state description. Different methods or cores are implemented in SOQCS to calculate the probability amplitudes and other characteristics at the circuit output. 

In the time span of SOQCS development other similarly focused software packages have arisen reflecting the interest in these kinds of simulations. For example, these include Perceval \cite{Perceval} and PhotoniQLAB \cite{PQLAb} libraries. Compared with those platforms SOQCS has a strong focus in simulating the effect of various circuit imperfections while still being friendly to users with limited programming expertise or physical knowledge. Previous libraries either do not support imperfections or they do it with less depth than SOQCS in a manner that requires an additional coding effort.

The paper is organized as follows. In section \ref{Intro} we will introduce some basic concepts and the overall architecture of the SOQCS library. In sections \ref{Architecture} and \ref{Structures}, we will explain the basic data structures and how they are used to perform a simulation. In section \ref{Simulator}, different simulation methods available in SOQCS are reviewed. In section \ref{Imperfections}, we explain some of the main physical imperfections implemented in SOQCS, such as partial photon distinguishability,  presence of mixed photonic states in the emission process and photon losses. Finally, in section \ref{Detectors} we will discuss how quantum measurement is treated in SOQCS. This includes the physical effects in the detection process and provides also details on how post-selection is defined and what data structures are needed to analyze the output.
The last section concludes the paper with a brief summary and outlook.

\section{Library architecture}
\label{Architecture}

\begin{figure}[h]
\includegraphics[width=0.48\textwidth]{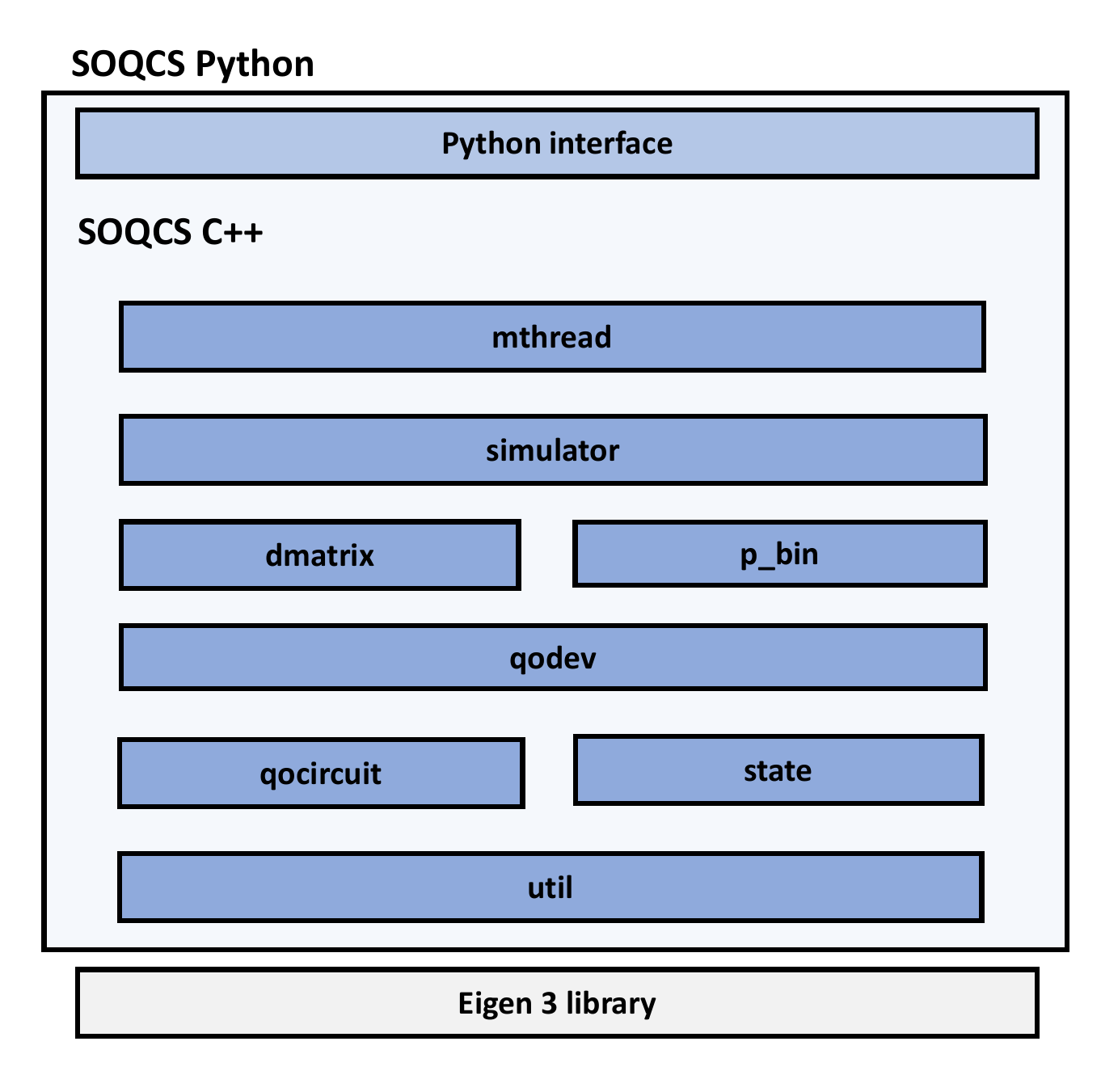}
\caption{ Diagram of the software structure of the SOQCS library. 
SOQCS is built in layers on top of Eigen3 \cite{Eigen3} library for linear algebra, with the level of abstraction increasing upwards.
Each layer contains C++ classes  that use other more basic classes defined on lower layers.}
\label{F1}
\end{figure}

SOQCS library is built with the objective to simulate general quantum optical circuits where imperfections are present. We define a model for an optical circuit from its interconnected components characterized by a set of suitable parameters.  These can be configured to fit particular simulation objectives that may include imperfections. We are particularly interested in a modular approach on how circuits are built and configured by a library user. Moreover we are also interested in the modularity of the library architecture itself, so simulations can be easily adapted to different objectives.

Upon configuration, a user may choose from a catalog of devices and imperfections. The library is built in a way that allows these catalogs to grow and be integrated with the rest of the library in the easiest way possible. This modular philosophy is behind many of the design decisions of SOQCS and will be revisited throughout the paper.

For this reason, SOQCS has been conceived as a stack of layers built on top of the  Eigen3 C++ template library for linear algebra and matrix manipulation \cite{Eigen3} (see fig. \ref{F1}). Each layer contains definitions of C++ classes that use other classes defined on lower layers while at the same time they are used by classes on higher layers. At the bottom of the library stack the custom made \verb|util| (utilities) sub-library resides which contains mathematical and dictionary management tools to perform multiple basic operations all across the software stack.  

In the next layer, above \verb|util|, the C++ classes \verb|state| and \verb|qocircuit| are found. These classes represent a photonic quantum state and a quantum optical circuit respectively. In general, a state will be represented as a list of photon occupations by mode. Those modes represent  the channel, polarization and wavepacket that a photon may take in the circuit. On the other hand, the circuit class contains the matrix definition of the linear optical circuit and a dictionary that relates mode numbers with their meaning. Additionally, the circuit object contains methods to define its individual elements. Linear optical elements are represented internally by matrices while non-linear elements like delays or detectors are defined by virtual circuit elements that inform and drive the simulator actions (more on this in the next sections).  

Furthermore, \verb|qocircuit| also handles the definition of wavepackets (equivalent to single particle wavefunctions). In SOQCS those wavepackets are limited to Gaussian or Exponential shapes parametrized by real numbers. There is an infinite number of those wavepackets in principle. However, only a discrete set is used in each simulation to make the problem manageable by a computer with finite memory. 

The next layer contains the C++ class \verb|qodev| (or "quantum device"). The different C++ classes that describe states, circuits and packets are integrated into a single class to allow for a more transparent and easy to use configuration of a simulation. Sometimes we will refer to this layer as the abstraction layer because it creates an interface to define circuits, packets and detectors in a more human friendly form
than some of the classes it encompasses. 

On top of \verb|qodev|  are defined the classes \verb|dmatrix| and   \verb|p_bin| to handle the representation of mixed states and measurement outcomes in the form of density matrices and probability distributions respectively. Finally, the class \verb|simulator| defines the simulator that constitutes the core of the library. This class uses all the different data structures and methods defined below to transform an input state into an output one and to handle this result in an automated manner according to a set configuration.

At the top of the library we find the layer with the class  \verb|mthread| that allows the execution of various simulations in parallel to speed up for example the calculations of density matrices representing the output of a circuit. Finally, at the highest level, an interface that encapsulates the C++ library is located allowing a Python interpreter to call SOQCS as a Python library. This functionality has the double advantage of making SOQCS accessible to both C++ and Python users and also use a more efficient C++ implementation in the more widespread and flexible Python environment.

\section{Data structures}
\label{Structures}
The simulator transforms photonic input states of a circuit $|\vec{\Psi}_{in} \rangle$ into output states $|\vec{\Psi}_{out} \rangle$.  We restrict ourselves to optical circuits made of linear optical elements that can be represented by matrices \cite{Linear1,Linear2,KLM}. In the next sub-sections we discuss how states and linear elements are represented in SOQCS data structures. Additionally, we will show how linear elements are integrated into optical circuits.

\subsection{States}

The photonic states are defined as a linear superposition of kets where each ket in the basis describes a different configuration of the photon occupations by mode. That is,
\begin{equation}
|\Psi_{in}\rangle=\sum_i \alpha_i| n_{i,1}, n_{i,2}, ..., n_{i,n} \rangle\,,
\end{equation}
where $n_{i,j}\geq0$ is the occupation of mode $j$ in the ket $i$ and $\alpha_i$ is the relevant complex coefficient. A mode $j$ labels a combination of quantum numbers 
for the channel, polarization and other degrees of freedom. A state can be represented in a computer as a list of kets where each ket consists of a 2-tuple made of a complex amplitude and a list of photon occupations by mode,
\begin{equation}
\begin{split}
|\Psi\rangle&:=\sum_i \alpha_i |\phi_i \rangle = \sum_i \{\alpha_i, \vec{v}_i\}\\
&=\{\{\alpha_1,\vec{v}_1\},\{\alpha_2,\vec{v}_2\}...\{\alpha_n,\vec{v}_n\}\}=\vec{\Psi}\,.
\end{split}
\end{equation}
The simulator C++ class transforms the input state into the output state according to a circuit definition,
\begin{equation}
\sum_i \alpha_i |\phi_i \rangle \rightarrow \sum_i \alpha_i T(|\phi_i \rangle) \,,
\label{ETransform}
\end{equation}
where $T(|\phi_i \rangle)=\sum_j \beta_{i,j} |\phi_j \rangle = |\Psi_{aux} \rangle$ is a linear transformation.  

Note that the implementation of this transformation in the simulator class is oblivious of the meaning of each mode. We need to maintain a dictionary to keep track of the relationship between the labelling of the different modes and their meaning (channel, polarization, etc). This dictionary is stored as part of the circuit class, so it can be used by all states related to that circuit. The simulator itself only requires the input state and the matrix defining the circuit transformation to perform the calculation.  The circuit matrix is also stored as part of the circuit data structure.

SOQCS provides an interface to create, operate and display states. For example, in the snippet of code below,
\begin{verbatim}
# Create input state
inputst = 
soqcs.state(example.num_levels(),10)

# Initialize input state
# Add ket #1: |2,0>
term=[[0,1], # Channels
      [2,0]] # Occupations
inputst.add_term(1.0/sqrt(2.0),
                 term,example)

# Add ket #2: |1,1>
term=[[0,1], # Channels
      [1,1]] # Occupations
inputst.add_term(1.0/sqrt(2.0),
                 term,example)
\end{verbatim}
we can see how a state (\verb|inputst|) related to a circuit (\verb|example|) is initialized. The state $1/\sqrt 2(|2,0\rangle +|1,1\rangle)$ is created adding first the ket $|2,0\rangle$ and then the ket $|1,1\rangle$.

\subsection{Circuits}

\subsubsection{Basic concepts}
In SOQCS a quantum optical circuit is a C++ class. It contains both the dictionary that relates each mode index with its meaning (channel, polarization and wavepacket) and the transformation matrix \cite{Linear1,Linear2,KLM} that relates the input modes to the output ones.  

For example, if we consider a circuit made of a single beamsplitter parametrized by two angles ($\theta$ and $\phi$), then the  corresponding 
transformation matrix is,
\begin{equation}
	U=\begin{pmatrix}
	\cos(\theta) & - e^{i \phi}\sin(\theta)\\
	e^{-i \phi} \sin(\theta) & \cos(\theta)
	\end{pmatrix}\,.
\end{equation}
This matrix is encoding the relationship between input and output single photon creation operators,
\begin{equation}
	\begin{matrix}
		\hat{a}_1^\dagger \rightarrow & \cos(\theta)\hat{a}_1^\dagger + e^{-i \phi}\sin(\theta) \hat{a}_2^\dagger  \\
		\hat{a}_2^\dagger \rightarrow & -e^{i \phi} \sin(\theta)\hat{a}_1^\dagger + \cos(\theta) \hat{a}_2^\dagger,  
	\end{matrix}
\end{equation}
\\
\noindent This transformation can be used to calculate an output state of this simple circuit from any given input
state,
\begin{equation}
\begin{split}
|n_1, n_2 \rangle &=\frac{\bigl(\hat{a}_{1}^{\dagger}\bigr)^{n_1}} {\sqrt{n_1!}}\frac{\bigl(\hat{a}_{2}^{\dagger}\bigr)^{n_2}}{\sqrt{n_1!}}|0, 0 \rangle \\
\rightarrow \frac{1}{\sqrt{n_1!n_2!}} &\Bigl( \hat{a}_{1}^{\dagger}\cos\theta +\hat{a}_{2}^{\dagger}e^{-i\phi}\sin\theta\Bigr)^{n_1} \\ 
& \Bigl(-\hat{a}_{1}^{\dagger}e^{i\phi}\sin\theta + \hat{a}_{2}^{\dagger}\cos\theta\Bigr)^{n_2}|0, 0 \rangle\,.
\end{split}
\end{equation}
For two photons inputs the corresponding outputs of this beamsplitter are,
\begin{small}
\begin{equation}
\begin{split}
|1, 0 \rangle  &\rightarrow \cos\theta |1, 0 \rangle  + e^{-i\phi}\sin\theta |0, 1 \rangle \\
|0, 1 \rangle  &\rightarrow -e^{i\phi}\sin\theta |1, 0 \rangle +  \cos\theta |0, 1 \rangle\\
|1, 1 \rangle  &\rightarrow -\sqrt{2}e^{i\phi}\cos\theta \sin\theta |2, 0 \rangle + (\cos^{2}\theta -\sin^{2}\theta)|1, 1 \rangle \\
&+\sqrt{2}e^{-i\phi}\cos\theta \sin\theta |0, 2 \rangle \\
|2, 0 \rangle  &\rightarrow \cos^{2}\theta |2, 0 \rangle + \sqrt{2}e^{-i\phi}\cos\theta \sin\theta |1, 1 \rangle + e^{-2i\phi}\sin^{2}\theta |0, 2 \rangle\\
|0, 2 \rangle  &\rightarrow e^{2i\phi}\sin^{2}\theta |2, 0 \rangle - \sqrt{2}e^{i\phi}\cos\theta \sin\theta |1, 1 \rangle + \cos^{2}\theta |0, 2 \rangle
\end{split}
\end{equation} 
\end{small}
To obtain an output state from an input state given any general circuit it is necessary to:
\begin{itemize}
\item Describe the circuit providing a list of individual elements and their interconnections.
\item Build the circuit transformation matrix from that description.
\item Apply a method to perform the transformation operation using the circuit matrix (and the input state).
\end{itemize}
The two first points will be covered in the rest of this section while the last point is the purpose of the simulator class and it will be treated in the next section. 

Naturally, a transformation matrix for each individual optical element has to be available in the library. The transformation matrices of many different linear optical elements can be found in the literature \cite{Linear1,Linear2,KLM}. The list of these elements implemented in SOQCS is provided in appendix A. Note that some of them, like detectors or delays perform non-linear operations that are treated in a different way than the linear elements explained here. We will discuss those below.

\subsubsection{Circuit matrix construction}

The circuit transformation requires a matrix $U$ that maps input into output creation operators for a particular optical circuit. The operation of building the matrix $U$ from an enumeration of individual optical elements is divided in two steps: embedding and composition. 
\begin{itemize}
\item {\bf Embedding}: An individual optical element only connects a subset of the available modes in the full circuit. Here we must identify the relationship between the local definition of an optical element and the degrees of freedom of the entire circuit. For example, we may recall that the matrix representation of a beamsplitter is given as 
\begin{equation}
B_{AB}=\begin{pmatrix}
\cos(\theta) & - e^{i \phi}\sin(\theta)   \\
e^{-i \phi} \sin(\theta)   &  \cos(\theta)
\end{pmatrix}
=\begin{pmatrix}
a & b   \\
c & d
\end{pmatrix}\
\end{equation}
where each row and column refers to the channels $A$ and $B$. If we assume a four channel circuit (without additional considerations about polarization or other degrees of freedom) then
we have to specify to which channels the beamsplitter is connected. If $A=2$ and $B=3$ then,
\begin{equation}
B_{23}=
\begin{pmatrix}
1 & 0 & 0 & 0 \\
0 & a & b & 0 \\
0 & c & d & 0 \\
0 & 0 & 0 & 1 
\end{pmatrix}\,.
\end{equation}		
Embedding consists of replacing the appropriate submatrix of the identity matrix by the matrix representation of the optical element.
	
\item {\bf Composition}: The total circuit matrix $U$ is the product $U=U_n \cdot... \cdot U_2 \cdot U_1$ where $U_1, U_2, ... U_n$ are the matrices of each of the individual optical elements.
Note that, here we are neglecting effects of backscattering. Additionally, we assume short photon packets in relation with the distance between optical elements
thus also neglecting resonance effects due standing waves.
\end{itemize}

\begin{figure}[h]
\includegraphics[width=0.5\textwidth]{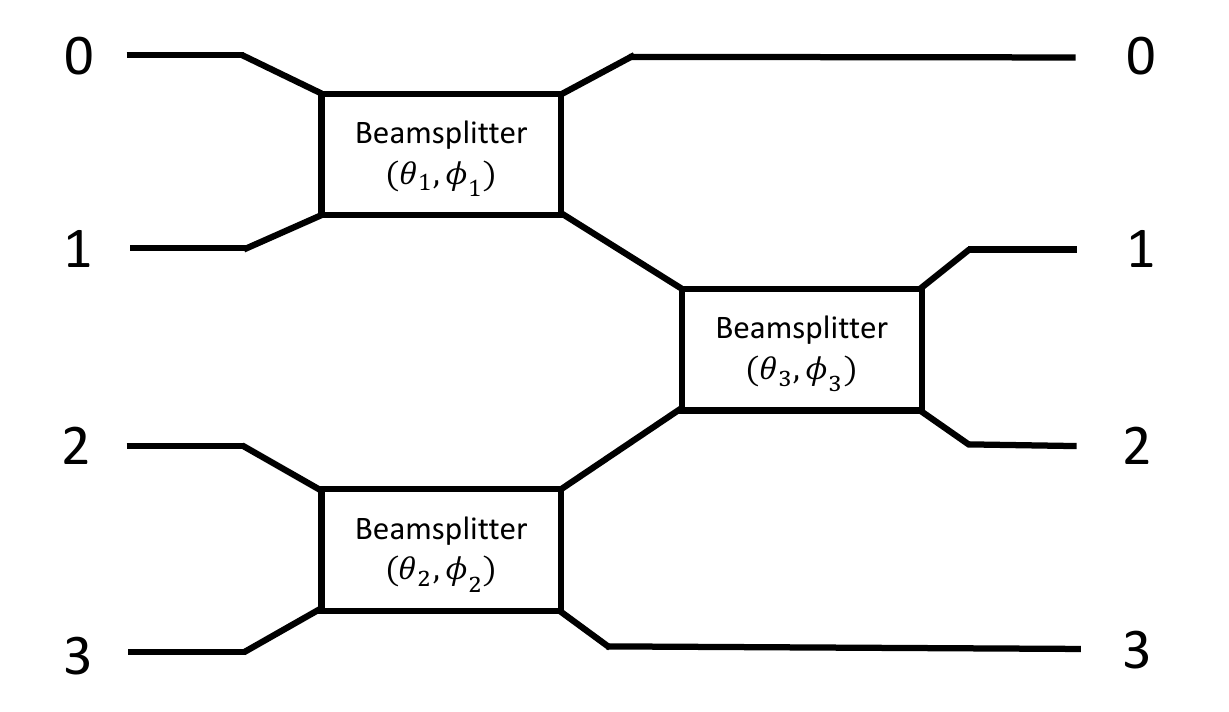}
\caption{Diagram of an arbitrary optical circuit made of three beamsplitters.}
\label{F3}
\end{figure}

As an example, the circuit depicted in fig \ref{F3} can be described in SOQCS by the set of instructions,
\begin{verbatim}
circuit.beasmplitter(0,1,theta1,phi1);
circuit.beasmplitter(2,3,theta2,phi2);
circuit.beasmplitter(1,2,theta3,phi3);
\end{verbatim}
that will translate into the matrix $U=U_{12}U_{23}U_{01}$ where $U_{ij}$ is a beamsplitter matrix for channels $i$ and $j$. In this set of instructions we list the components of the circuit, their connections but
also the parameters of each of the beamsplitters. 

Overall, SOQCS may represent any optical linear circuit just by listing its elements (with their corresponding parameters) and the connections between them. The construction of the entire circuit matrix is performed in an automated manner.

\begin{figure}[h]
\includegraphics[width=0.5\textwidth]{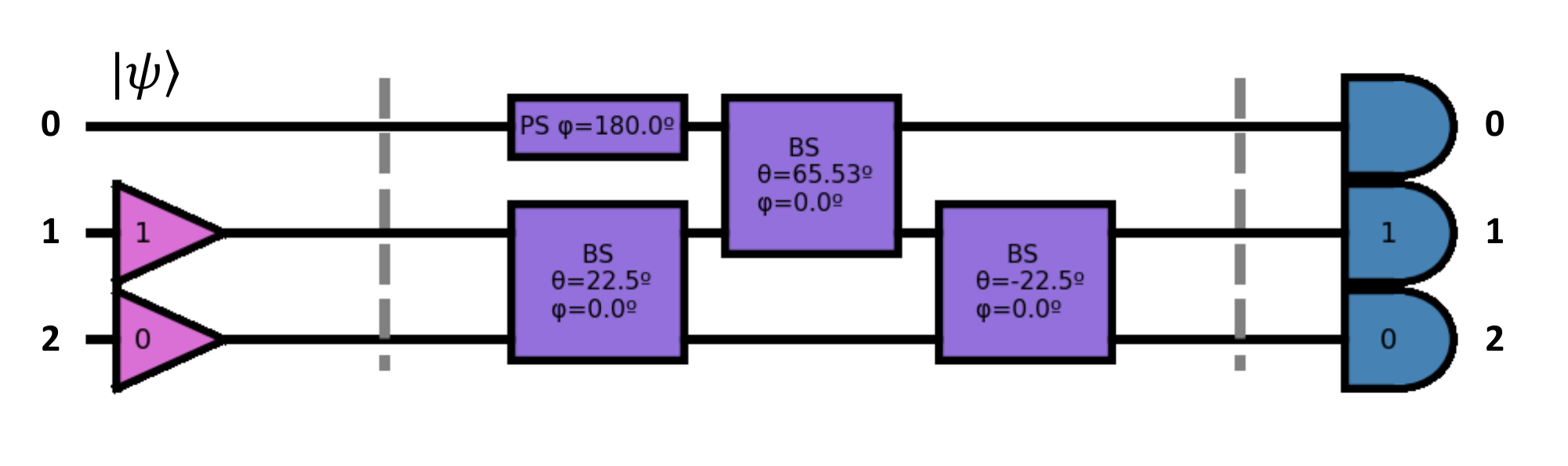}
\caption{NSX circuit diagram as discussed in ref. \cite{Linear1}. Except for the ket $|\Psi\rangle$ symbol this figure has been created by SOQCS graphical module. Emitters are plotted as triangles while detectors
are implemented as square like shapes with one round side. The optical elements are represented as boxes. PS stands for phase shifter while BS for beamsplitter.}
\label{F4}
\end{figure}

\subsubsection{Example: NSX circuit}
We present as an example how a NSX circuit \cite{Linear1}, illustrated in fig.\ref{F4}, is created and used in a simulation. First, the circuit is created adding each individual optical element,
\begin{verbatim}
# Create circuit
circuit = soqcs.qocircuit(3);

# Build circuit
circuit.phase_shifter(0,180.0);
circuit.beamsplitter(1,2,22.5,0.0);
circuit.beamsplitter(0,1,65.5302,0.0);
circuit.beamsplitter(1,2,-22.5,0.0);
circuit.detector(0)
circuit.detector(1,1)
circuit.detector(2,0)
\end{verbatim}
then the initial state of the circuit is created. In this case we will select as input the bosonic state $|\Psi\rangle=|0\rangle+|1\rangle+|2\rangle$ in channel zero while an ancilla photon is added to channel one.

\begin{verbatim}
# Create input state
inputst = 
soqcs.state(circuit.num_levels(),10)

# Adds ket |0>
term=[[0,1,2], # Channels
      [0,1,0]] # Occupations
inputst.add_term(1.0,term,circuit)

# Adds ket |1>
term=[[0,1,2], # Channels
      [1,1,0]] # Occupations
inputst.add_term(1.0,term,circuit)

# Adds ket |2>
term=[[0,1,2], # Channels
      [2,1,0]] # Occupations
inputst.add_term(1.0,term,circuit)
\end{verbatim}
To perform the calculation it is needed to create and run a simulator C++ object:
\begin{verbatim}
simulator=soqcs.simulator()
outputst=simulator.run_st(inputst,circuit)
\end{verbatim}
We will cover the inner workings of the \verb|simulator| class in the next section. 
Finally, the output is post selected by the condition $|1,0\rangle$ in channels one and two. This condition is specified above as a part of the detectors definition in the circuit,
\begin{verbatim}
outputst=circuit.apply_condition(outputst)
\end{verbatim}
The final state is printed using the command,
\begin{verbatim}
outputst.prnt_state(column=1)
\end{verbatim}
which provides the intended output state with the phase flip operation applied to the term with two photons,
\begin{verbatim}
| 0 >:  0.49999999 + 0.00000000 j
| 1 >:  0.50000001 - 0.00000000 j
| 2 >: -0.50000000 + 0.00000000 j
\end{verbatim}
Note that the state is intentionally not normalized to unity in order to provide information about the probability of success of post-selection which is $P=0.25$.  This is not the only way to perform post-selection in SOQCS. Any projector can be defined in a manner similar to definition of states in order to perform the post-selection operation.

In appendices B and C, examples for the CZ \cite{KLM} and CNOT \cite{CNOT} gates are provided. These examples present different ways to perform simulations and also show how SOQCS can be used to encode qubits into photonic states.

\section{Simulator core}
\label{Simulator}
As explained above, the \verb|simulator| C++ class is used to transform input states into output states. The eq. \ref{ETransform} that describes this transformation can be  written in algorithmic form,

\begin{algorithmic}
\State $\vec{\Psi}_{out} = \{\}$
\For{i=0 to i=(size of $\vec{\Psi}_{in}$)} 
\State $\vec{\Psi}_{in}(i) \rightarrow \{\alpha_i,\vec{\phi}_i\}$
\State $T(\vec{\phi}_i) \rightarrow \vec{\Psi}_{aux}$
\For{j=0 to j=(size of $\vec{\Psi}_{aux}$)} 
\State $\vec{\Psi}_{aux}(j) \rightarrow \{\beta_j,\vec{\psi}_j\}$
\State $\vec{\Psi}_{out} \leftarrow$ add $\{\alpha_i \cdot \beta_j, \vec{\psi}_j\}$
\EndFor
\EndFor
\end{algorithmic}
where $\vec{\Psi}_{in}$, $\vec{\Psi}_{out}$, and $\vec{\Psi}_{aux}$ are the objects representing the input, output and auxiliary states as list of kets in the memory of a computer.  Note that zero amplitude kets do not need to be stored. 

The transformation of the input state is made ket by ket and each ket transformation $T(\vec{\phi}_i,U)$ depends on the circuit matrix $U$.  Note that, the sum of two states "add"  is equivalent to the concatenation of two lists of kets provided that no ket is repeated. In the case of a repetition the amplitudes of those duplicated are summed. We store a dictionary  as part of the state data structure to check for repetitions in a straightforward manner.

The function $T(\vec{\phi}_i,U)$ defines the core operation of the simulator and can be implemented using different methods. SOQCS implements several of them and provides an interface that allows for new ones to be added in the future. The algorithms implementing those methods have a common structure, 
\begin{algorithmic}
\State $\vec{\Psi}_{aux} = \{\}$
\For{$\forall \vec{\psi}_j$ in $\mathcal B$} 
\State $\beta_j=f(\vec{\phi}_i,\vec{\psi}_{j},U)$
\State $\vec{\Psi}_{aux} \leftarrow$ add $\{\beta_j, \vec{\psi}_j\}$
\EndFor
\end{algorithmic}
where a probability amplitude $\beta_j$ is calculated for each possible output ket $\vec{\psi}_{j}$ (given an input one $\vec{\phi}_i$ and the circuit matrix $U$). The transformation $f$ is different to each chosen method.
The output kets are taken from an ensemble $\mathcal B$ made of an orthonormal subset of kets relevant for the current simulation. At this moment in SOQCS there are implemented two methods to perform this operation and these methods may be used in three different ways depending on $\mathcal B$.
\begin{itemize}
	\item  $\mathcal B$ is made of kets with the same number of photons as the input ket. In this case, the full output distribution is calculated.
	\item   $\mathcal B$ also assumes photon number conservation but is restricted to include only kets with a maximum of one photon per mode. The overall computation time with respect of the previous option is improved because   we are restricting ourselves to a smaller Hilbert subspace. This is useful for circuits implementing qubit operations because the usual encodings of qubits into photonic states do not involve more than one photon by channel.  
	\item  $\mathcal B$ is chosen specifically by the user. Only  outcomes that are relevant for a particular simulation are calculated. Those that are considered unimportant are neglected. For example, in the calculation of the fidelity of a particular gate with respect to some parameters only those kets involved in the ideal solution are of interest. Restricting the calculation to even a smaller subset of the Hilbert space  makes the overall algorithm faster compared to the previous options.
\end{itemize}

Next, we review the algorithms which implement $T(\vec{\phi}_i,U)$  and which are part of SOQCS.

\subsection{Direct method}
The most straightforward way to perform the single ket transformation $T(\vec{\phi}_i,U)$  is to imitate how the analytical calculation is performed. First, it is convenient to re-interpret the circuit matrix transformation as just a list of rules of the form,
\begin{equation}
a^\dagger_i \rightarrow U_{1,i} a^\dagger_1 + U_{2,i} a^\dagger_2 + ... + U_{n,i} a^\dagger_n
\end{equation}
where $a^\dagger_i$ is the creation operator of a photon in the mode $i$ and $U_{j,i}$ is the  probability amplitude of finding a photon in the output mode $j$ if there is one present in the input mode $i$. Therefore, for a given
input occupation vector $\vec{\phi}_i=\{n_1,n_2,...,n_l\}$,
\begin{strip}
\begin{equation}
\begin{split}
&(a^\dagger_1)^{n_1}(a^\dagger_1)^{n_2}...(a^\dagger_1)^{n_l} \rightarrow \left(\sum_j^l U_{j,1} a^\dagger_j \right)^{n_1} \left(\sum_j^l U_{j,2} a^\dagger_j\right)^{n_2}...\left(\sum_j^l U_{j,l} a^\dagger_j\right)^{n_l}= \\
&=\sum_{\forall \vec{p}} \underbrace{U_{p_1,1}\cdot U_{p_2,1}\cdot ...\cdot U_{p_{n_1},1}}_\text{Repeated $n_1$ times}\underbrace{U_{p_{n_1+1},2}\cdot U_{p_{n_1+2},2}\cdot ... \cdot U_{p_{n_1+n_2},2}}_\text{Repeated $n_2$ times} ... \cdot U_{p_{N},l} a_{p_1}^\dagger \cdot a_{p_2}^\dagger \cdot ...\cdot a_{p_{N}}^\dagger=\\
&=\sum_{\forall \vec{p}} \alpha_{\vec{p}}\,\, a_{p_1}^\dagger \cdot a_{p_2}^\dagger \cdot ...\cdot a_{p_{n}}^\dagger=\sum_i  \alpha_i |\phi_i\rangle
\end{split}
\end{equation}
\end{strip}
where $l$ is the number of modes, $\vec{p}$ is and index list $\vec{p}=\{p_1,p_2,...,p_n\}$ and  $n=n_1+n_2+...+n_l$ is the total number of photons.  Thus, the simulation consists of generating all the possible sequences of  indexes $\vec{p}$ from $1,1,..1$ to $l,l,...l$ that belong to $\mathcal{B}$. Note that different sequences of indexes may lead to the same photon occupation by mode.

This straightforward method scales quite poorly. To calculate a single probability amplitude of the output it is needed to consider $n!/n_1!n_2!..n_n!$ sequences of indexes. Each sequence implies the multiplication of $n$ coefficients to obtain the probability amplitude. This calculation can be optimized if zeros are present in the multiplication. In a worst case scenario, the computational cost scales as $O(n n!)$ . In spite of its unfavorable scaling properties, it offers the best cost for problems with a small number of photons due its simplicity.

\subsection{Permanent method}
A known approach to obtain the probability amplitude of a photonic circuit outcome is based on the calculation of the permanent \cite{Permanent} of a square matrix. This matrix is built from the same coefficients as the circuit matrix but with a different number of rows and columns. If the amplitude of an output ket is calculated given an input one, columns that correspond to the transformation of a determined mode are repeated as many times as that mode is occupied in the input ket. On the other hand, rows are repeated as many times as their corresponding mode is occupied in the output ket. Various algorithms are available in the literature to perform a permanent calculation \cite{Ryser,Glynn}. In SOQCS we calculate permanents using Ryser \cite{Permanent2} 
\begin{equation}
perm(A)=(-1)^n \sum_{S\subseteq\{1,...,n\}} (-1)^{|S|} \prod_{i=1}^n \sum_{j \in S} a_{i,j}
\end{equation}
and the Balasubramanian-Bax-Franklin-Glynn \cite{Glynn} formulas  implemented in gray code, 
\begin{equation}
perm(A)=\frac{1}{2^{n-1}}\left[\sum_\delta \left( \prod_{k=1}^n \delta_k \right)\prod_{j=1}^n \sum_{i=1}^{n}\delta_i a_{i,j}  \right]
\end{equation}
where for each amplitude the computational cost is $O(n2^n)$ where $n$ is the number of photons.

\subsection{Benchmarks}
Here we present benchmark results of SOQCS for single thread simulations performed on an Intel I7-10750H@2.60GHz with 16GB of RAM memory. These benchmarks have been carried out using
circuits represented by matrices of random coefficients. These circuits are initialized with states that do not contain more than one photon by channel. The first benchmark consists in measuring the time 
that takes to calculate an output state.

\begin{center}
\begin{table*}[h]
\begin{small}
\begin{center}
\begin{tabular}{  |c|c|c|c|c|c| } 
\hline
Photons       &Channels      &Direct F(ms)  &Direct R(ms)  &Glynn F(ms)   &Glynn R(ms) \\
\hline
2             &4             &3             &3             &3             &3 \\
3             &6             &3             &3             &3             &3 \\
4             &8             &3             &3             &3             &3 \\
5             &10            &19            &6             &7             &4 \\
6             &12            &544           &69            &37            &6 \\
7             &14            &18040         &1576          &346           &18\\
\hline
\end{tabular}
\end{center}
\end{small}
\caption{Calculation time of each method implemented in  SOQCS to obtain the output state of a random circuit.}
\label{Table_1}
\end{table*}
\end{center}

The results are presented in table \ref{Table_1} where in the \verb|Direct F| and \verb|Glynn F| columns are shown the computation times for the straightforward and permanent methods respectively. These methods calculate the amplitude distribution of the output state considering a base of kets with the same number of photons as the input. On the other hand, \verb|Direct R| and \verb|Glynn R| are their "restricted" counterparts where only outputs with a maximum of one photon per mode are considered. The smaller the Hilbert space of  possible solutions the faster the result is obtained. Both the direct and the permanent method perform similarly up to four photons. Permanent methods scale better for a large number of photons.

\begin{table}[htbp]
\begin{small}
\begin{tabular}{  |c|c|c|c| } 
\hline
Photons   &Channels   &Direct(ms)   &Glynn(ms)\\
\hline
2         &4          &0            &0\\ 
4         &8          &0            &0\\    
6         &12         &0            &0\\    
8         &16         &20           &0\\    
9         &18         &38           &0\\    
10        &20         &393          &0\\    
20        &40         &-            &113\\  
22        &44         &-            &337\\  
24        &48         &-            &1220\\ 
26        &52         &-            &5157\\ 
28        &56         &-            &21864\\
30        &60         &-            &92386\\
\hline
\end{tabular}
\end{small}
\caption{Calculation time of each method implemented in  SOQCS to obtain the probability amplitude of a single outcome. }
\label{Table_2}
\end{table}

Times needed to obtain a selected single probability amplitude are shown in table \ref{Table_2}. These results are obtained using similar random circuits and inputs as in the previous benchmarks.

It is feasible to calculate in a reasonable time a single probability amplitude for circuits up to 30 photons.  This limit can be improved using parallelism in the permanent calculation. 
However, the calculation of an output state implies the calculation of the full distribution of probability amplitudes. This imposes a more realistic limit of 7 to 9 photons independently of the use of parallelism.
The overhead of performing a parallel calculation is usually unfavorable for this range of photons. Therefore we prefer to parallelize the calculation of different output states. This is useful when calculating output density matrices.

\subsection{Sampling}
Sampling means generating instances from the probability distribution of the output of the circuit. In SOQCS, it is implemented using the Clifford A algorithm described in ref. \cite{Sampling} and a metropolis algorithm presented in ref. \cite{Sampling2}. 

\subsubsection{Clifford algorithm}
The Clifford A algorithm \cite{Sampling} relies on the fact that the probability to measure $n$ photons in a particular set of modes depends on a chain of conditional probabilities,
\begin{equation}
\begin{split}
&p(r_1,r_2,...,r_n)=\\
&p(r_1)p(r_2|r_1)...p(r_n|r_{n-1},...r_{2},r_{1})\,,
\end{split}
\end{equation}
where $r_i$ represents one of the $l$ available modes. 

The algorithm to obtain a single sample is structured as a loop. In each iteration of the loop the probability of a photon $i$ to be in a given mode is calculated assuming knowledge of the modes occupied by the previous $i-1$ photons. The mode corresponding to photon $i$ is chosen randomly from the probability distribution of that photon to be at each mode. These probabilities are also calculated using permanents.  We refer to ref. \cite{Sampling} for further details regarding this algorithm. The loop is finished when all photons have been allocated. This algorithm has a computational cost of $O(nl3^n)$. 

\subsubsection{Metropolis algorithm}
In the metropolis algorithm proposed in ref. \cite{Sampling2} successive samples $s$ with probability $P(s)$ are proposed with probability $Pc(s)$. The transition between these samples is  accepted with probability,
\begin{equation}
T(s'|s)=min\left(1,\frac{P(s')Pc(s)}{P(s)Pc(s')}  \right)
\end{equation}
where $s$ is the previous accepted sample, $s'$ the new sample proposed and $Pc(s)$ is taken as the classical distribution where photons are distinguishable particles. The result is a Markov chain that in its stationary regime produces samples that approximate the desired distribution $P(s)$. 

This algorithm has a computational cost of $O(n2^n)$ (like Clifford B) and it needs to calculate only one permanent by sample. Note that a burn period and a thinning procedure may be needed to make the chain independent of the initial state and avoid sample correlations respectively. 

\subsubsection{Example: Sampling comparison with the exact result}
In the following example we perform a comparison between the probability distribution obtained by the direct method and the Clifford A sampler. The circuit is initialized in a way that provides probability distributions automatically in both cases.

\begin{verbatim}
# Build the circuit
example = soqcs.qodev(2,4)
example.add_photons(1,0)
example.add_photons(1,1)
example.random_circuit()
example.detector(0)
example.detector(1)
example.detector(2)
example.detector(3)

# Create the simulator 
sim = soqcs.simulator()
# Run a simulation
apdexact=sim.run(example)
# Run a sampling procedure 
# with one million samples.
apdsample=sim.sample(example,1000000) 

# Plot results
apdexact.show(dpi=70)
apdsample.show(dpi=70)
\end{verbatim}

\section{Imperfections}
\label{Imperfections}
Simulation methods are used to obtain output states or samples of ideal circuits. However, different kinds of imperfections are relevant to photon emission, photon transfer across
a circuit or to photon detection. Some circuit imperfections can be handled without considering any additional mechanism. For example, unbalanced beamsplitters will imply $\theta\neq 45^\circ$.
However, some other imperfections require additional support from the SOQCS library. 

In this section we will explain how SOQCS deals with three of the most common imperfections in optical circuits. That is, partial distinguishability of photons, imperfect sources and losses. Additional imperfections related to the detection process will be treated in section \ref{Detectors}.

\subsection{Partial distinguishability}
\subsubsection{Basic mechanism}
Partial distinguishability between photons is related to the partial overlap between their wavefunctions. If the overlap is total they are indistinguishable from each other. 
Partial distinguishability may happen for many reasons like small relative delays between photons or differences in their frequencies.

Using again as an example the beamsplitter, we can extend the definition of a ket as $|n_{ch=0,d=0},n_{ch=0,d=1},n_{ch=1,d=0},n_{ch=1,d=1}\rangle$ where $ch$ is the channel degree of freedom and $d$ is a label given to a particular wavepacket that characterizes the photons in that mode. If all photons are indistinguishable only modes that share the same $d$ label will be populated. This results in 
the following transformation:
\begin{equation}
|1, 0, 1, 0 \rangle  \rightarrow -\frac{1}{\sqrt{2}}|2, 0 , 0 , 0 \rangle +\frac{1}{\sqrt{2}} |0, 0, 2, 0 \rangle \,,
\end{equation}
where photons perfectly bunch in one of the output modes as explained by the Hong-Ou-Mandel (HOM) effect \cite{HOM}. On the other hand, if photons are distinguishable, inputs with different $d$ labels are populated.
This leads to an output,
\begin{equation}
\begin{split}
|1, 0, 0, 1 \rangle  \rightarrow &- 0.5\,| 1, 1, 0, 0 > - 0.5\,| 0, 1, 1, 0 > \\ 
&+ 0.5\,| 1, 0, 0, 1 > + 0.5\,| 0, 0, 1, 1 >
\end{split}
\end{equation}
where all possible outcomes have the same probability. Note that if photons are distinguishable the output probability distribution of a circuit can be calculated by classical means.

In the general case of partial distinguishability the wavepacket states are not orthogonal $\langle  P_i | P_j \rangle\neq 0$. They can be transformed into an orthonormal basis using the Gram-Schmidt orthonormalization procedure as suggested in ref. \cite{Distin}:
\begin{equation}
\begin{split}
|\tilde P_0 \rangle &= | P_0 \rangle ,  \\
|\tilde P_1 \rangle &= \frac{ | P_1\rangle -  | \tilde P_0 \rangle \langle \tilde P_0 | P_1 \rangle } {\sqrt{1-|\langle \tilde P_0 | P_1 \rangle|^2 }} , \\
|\tilde P_2 \rangle &= \frac{ | P_2 \rangle-  | \tilde P_0 \rangle \langle \tilde P_0 | P_2 \rangle - | \tilde P_1 \rangle \langle \tilde  P_1 | P_2 \rangle   } {\sqrt{1-|\langle \tilde P_0 | P_2 \rangle |^2- | \langle \tilde P_1 | P_2 \rangle|^2 }} \,. 
\end{split}
\label{EGram}
\end{equation}

A software implementation of partial distinguishability of photons is quite involved. We provide an extended discussion of this topic in ref. \cite{DistinSOQCS}. The key points are summarized below.

If photon wavepackets are included in the mode definition then the simulation methods presented above for ideal circuits can also be used for partially distinguishable photons. Photon wavepackets are treated on the same footing as the channel or polarization degrees of freedom. As a consequence, the implementation of the simulation methods is independent of the distinguishability model. This is an advantage because this independence allows for  both calculations to be developed separately as different reusable modules.

In SOQCS packets may be configured to be of Gaussian or Exponential shape. Gaussian packets,
\begin{equation}
| \omega_i \rangle =  \int dt 	\left( \frac{\sqrt{\Delta \omega_i}}{\pi^{1/4} } \right) e^{-(t-t_i)^2 \Delta \omega_i^2 } e^{-i \omega_i (t-t_i)} | t \rangle \,.
\end{equation}
are characterized by a central time $t_i$ a central frequency $\omega_i$ and width $\Delta \omega_i$ where $t$ is time. On the other hand, exponential packets,
\begin{equation}
| \omega_i \rangle =  \int dt \left(\frac{1}{\sqrt{\tau_{x_i}}} \right) \Theta (t-t_i)  e^{-\frac{t-t_i}{2 \tau_{x_i} } } e^{-i \omega_i (t-t_i)} | t \rangle \,,
\end{equation}
are characterized by a characteristic time $t_i$ and a decay time $\tau_{x_i}$. 

Physically, there is an infinite set of  wavepackets that represent the spatial degrees of freedom of the photons. To implement partial indistinguishability in SOQCS photons are restricted to a discrete finite subset. This discretization allows to define both an emitter and a delay optical gates in a similar manner as a linear optical element. This means that they are represented by their respective matrices. 

The emitter matrix contains the transformation to orthogonal packets derived from eq. \ref{EGram}, 
\begin{equation}
\begin{split}
| P_0 \rangle &\rightarrow | \tilde P_0 \rangle c_{0,0} \\
| P_1 \rangle &\rightarrow | \tilde P_0 \rangle c_{1,0} + |\tilde P_1 \rangle   c_{1,1}  \\
| P_2 \rangle &\rightarrow | \tilde P_0 \rangle c_{2,0} + |\tilde P_1 \rangle   c_{2,1}   + | \tilde P_2 \rangle c_{2,2} \,.
\end{split}
\end{equation}
On the other hand, time may be split into different periods where the same set of packets is defined in each period but displaced in time. Therefore, the orthonormalization of packets is also
the same at those different periods. We assume that the time between periods is large in comparison with packets width. A delay of one photon by one period of time implies the substitution of its packet label for the equivalent label in the next period, 
\begin{equation}
\begin{split}
\hat{T}_D =\sum_{j=0}^{n_t} | \tilde P_{j+n_t} \rangle  \langle \tilde P_j |\,.
\end{split}
\end{equation}
The delay matrix reflects the relabelling of packets at different discrete periods. 

Although those matrices are non-unitary these elements can be handled in the same way as any other linear optical element. The reason is that in this kind of simulation optical circuits are interpreted as one way operations from the input to the output. 
For further details see ref. \cite{DistinSOQCS}.

\subsubsection{Photon as an abstraction}
\begin{figure}[h]
\centering
\includegraphics[width=0.35\textwidth]{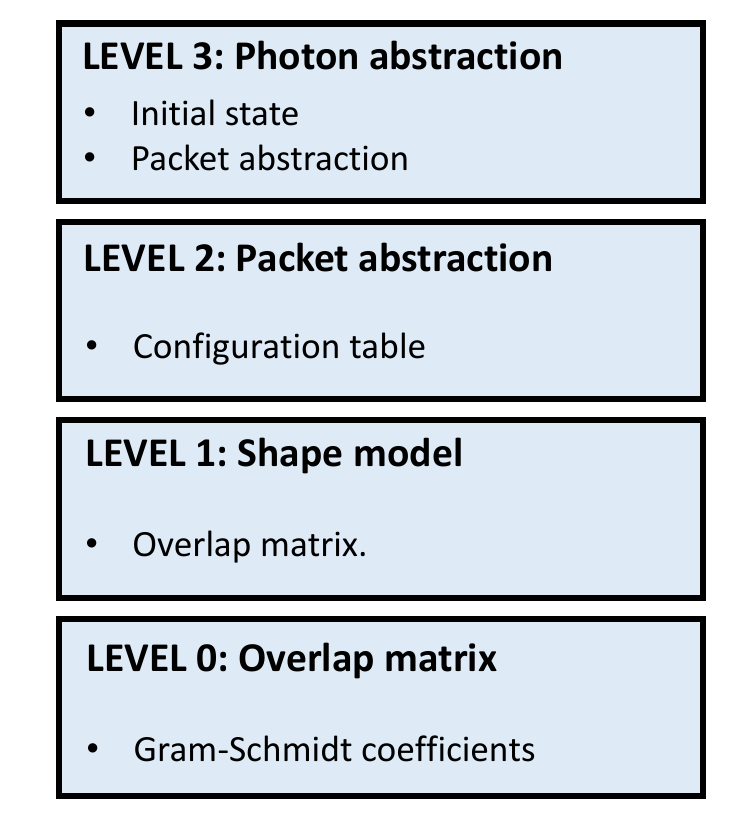}
\caption{Stack of abstraction levels used to implement partial distinguishability of photons in SOQCS library.}
\label{F5}
\end{figure}

To calculate partial distinguishability effects in SOQCS the Gram-Schmidt coefficients have to be known. 
However, the user would likely find cumbersome to define those coefficients directly or to create a code to do so. 
Instead, SOQCS only requires the user to define how many photons exist in each of the channels and which are the parameters given to those photons. 
To allow for this, we have implemented a stack of abstraction levels with the photon abstraction at the top as shown in fig. \ref{F5} and the  emitter and the delay definitions given by the Gram-Schmidt coefficients 
at the bottom. Each of the levels is responsible in organizing the information and performing the calculations needed to fill the gap between the two equivalent sets of information. The photon description
is understandable for a user while the Gram-Schmidt coefficients are the information needed by the computer to perform the actual calculation.

These levels are (in a brief description),
\begin{itemize}
\item Level 3: Photon abstraction level.  This is the upper abstraction level where photons are described by their paramenters (channel, time, frequency, etc). This information is splitted into the definition of the initial state of the circuit and a list of packets with their parameters.
\item Level 2: Packet abstraction level. In this level packets are treated as virtual circuit elements used to create a configuration table with the information of all packets. 
\item Level 1: Shape model. In this level the information stored in the configuration table is used to calculate the overlap matrix between packets. 
\item Level 0: Finally, in the bottom level the overlap coefficients are used to perform the Gram-Schmidt orthonormalization procedure. As a result, the emitter matrix to perform the transformation from the non-orthonormal basis to the orthonormal one is obtained.
\end{itemize}

The photon abstraction level is implemented in SOQCS as part of the device abstraction. The class \verb|qodev| is defined as an ensemble of an input state, a quantum optical circuit and a list of packets.
In this context, photons can be defined as virtual elements of the quatum circuit in a natural way without the need of defining separately the initial state and the packets related to the circuit. 
These elements are declared in the same way as the linear optical elements with the difference that they serve the purpose of configuring the simulation and have no matrix representation attached.
The class \verb|qocircuit| implements the circuit itself at a more fundamental level. In this class packets may be added to the circuit definition directly (level 2) also as virtual elements. 

In general, the function of the \verb|qodev| class is to allow for a more direct definition of photons. However, it also automates some aspects of the measurement process configuration. We will expand this matter further in section \ref{Detectors}.

\subsubsection{Example: Partial distinguishability of photons}
Next, we present the code for a circuit simulation where two groups of five photons are set up in each of the input channels of a  beamsplitter. These two groups of photons arrive to the beamsplitter with a relative delay $\Delta t$. The wavefunctions of both sets of photons are modeled as Gaussians of frequency $\omega=1$ and uncertainty $\Delta \omega=1$ in arbitrary adimensional units. The code is written in a function called \verb|HOMP| that returns the probability of the outcome specified by the parameter \verb|args| for a delay $\Delta t$ given by the parameter \verb|dt|. Detectors are configured as counters that measure the total number of photons independently of their detection time. Outcome probabilities are calculated automatically from the amplitudes of the output state. We will expand on how detectors are implemented in SOQCS in section \ref{Detectors}.

\begin{small}
\begin{verbatim}
def HOMP(dt,args):
    #Build the circuit
    example = soqcs.qodev(10,2,1,2,
                    0,10000,True);                
    example.add_photons(5, 0, 0, 
                 0.0, 1.0, 1.0);
    example.add_photons(5, 1, 0,
                  dt, 1.0, 1.0);
    example.beamsplitter(0,1,45.0,0.0);
    example.detector(0);
    example.detector(1);

    # Create a simulator and run the simulation
    sim=soqcs.simulator()
    measured=sim.run(example)    
                    
    # Return the probability
    och0=5+args[0]
    och1=5-args[0]
    term=[[ 0   ,    1],
          [ 0   ,    0],
          [ 0   ,    0],
          [ och0, och1]]
    prob=measured.prob(term,example)    
    
    return prob
\end{verbatim}
\end{small}

\begin{figure}[h]
\includegraphics[width=0.45\textwidth]{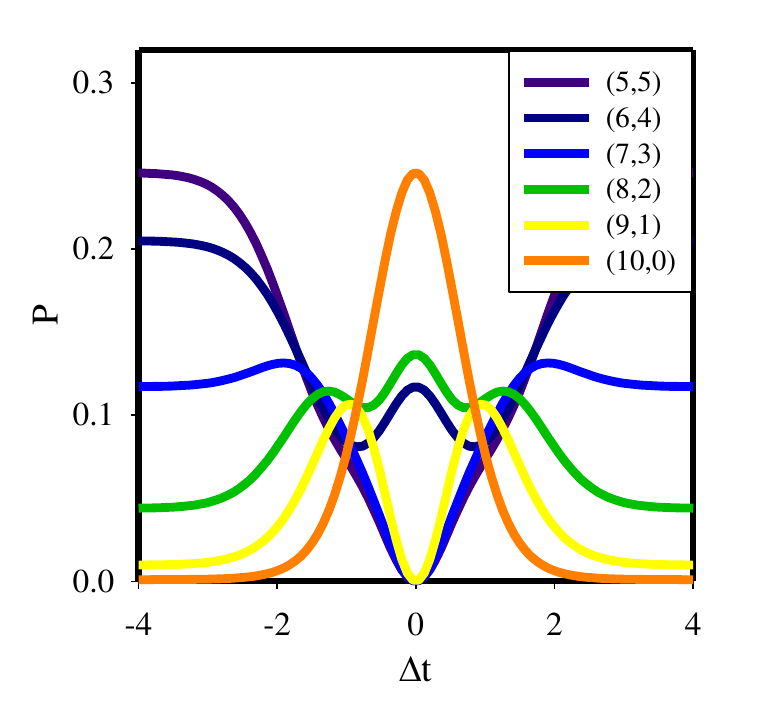}
\caption{ Variation of the probability with respect to $\Delta t$  of the different outcomes of a beamsplitter circuit where two groups of five photons of Gaussian shape are defined in each channel with a $\Delta t$ delay between them.}
\label{F6}
\end{figure}

Different outcome probabilities for each delay time $\Delta t$ are calculated by means of repeated calls to \verb|HOMP| and plotted in fig. \ref{F6}. This example is similar, but with more photons,
to examples discussed in \cite{Distin,DistinSOQCS}.

\subsection{Sources}
Physical photon emitters have imperfections. Therefore, emitted photons have to be represented by a density matrix instead of a pure state. In particular, SOQCS implements a parametrized quantum dot able to simulate the creation of pairs of entangled photons of features compatible with those emitted from a physical bi-exciton XX exciton X cascade.

\subsubsection{Quantum dot emission model}
From \cite{Coherence,Coherence2,Entanglement} the density matrix of photons emitted from a quantum dot cascade is,
\begin{equation}
\begin{split}
\hat{\rho}=&\left( 1-k\cdot p_s \right) \cdot\hat{\rho}_{noise}\,+\\ 
     &\left( k\cdot p_s \cdot (1-p_d) \right) \cdot \hat{\rho}_{pd}\,+\\
     &\left( k \cdot p_s \cdot p_d \right) \cdot \hat{\rho}_{en}\,.
\end{split}
\end{equation}
This matrix is written as the weighted sum of pure states density matrices. Each of these pure state density matrices are related with a physical process in the quantum dot. Here, $\rho_{en}$ is the density matrix of the entangled photons emitted in the ideal bi-exciton and exciton cascade. $\rho_{pd}$ is the density matrix of the photons where some cross-dephasing event played a role in their emission and finally,  $\rho_{noise}$ is the density matrix created by random photons from external sources of noise or photons emitted through any non-coherent process. The probability of each outcome is a function of $k$ that is the fraction of photons emitted by the quantum dot source. On the other hand, $(1-k)$  is the fraction caused by noise. $p_s=e^{-\frac{t}{t_{ss}}}$ is the probability of the dot to emit coherently, $t_{ss}$ being the characteristic time of coherence. Finally, $1-p_d=1-e^{-\frac{t}{t_{HV}}}$ is the probability of a dephasing event where $t_{HV}$ is the cross dephasing characteristic time.

Here is where the stochastic nature of the simulation plays an important role. 
To calculate the output density matrix of a circuit, the possible input pure states are sampled according to their corresponding probabilities and simulated individually. The output states are then used to calculate the resulting density matrix. The main advantage of this scheme is that it can be implemented using computation methods developed for pure states and ideal circuits while
achieving our objective of modularity.

The density matrix of the entangled photons  $\rho_{en}$ is given by,
\begin{equation}
\rho_{en}=\frac{1}{2}\left(\begin{matrix}
1          & 0 & 0 & e^{-iS\Delta t/h} \\
0          & 0 & 0 & 0          \\
0          & 0 & 0 & 0          \\
e^{+iS\Delta t/h} & 0 & 0 & 1 
\end{matrix}
\right)\,,
\end{equation}
in the basis $\{|HH\rangle, |HV \rangle , |VH \rangle, |VV \rangle\}$ where $H$ and $V$ are the "horizontal" and "vertical" polarizations respectively. The density matrix above represents the pure state, 
\begin{equation}
|\Psi\rangle=|HH\rangle +  e^{-iS\Delta t/\hbar} | VV \rangle\,.
\end{equation}
where  $S$ is the fine structure splitting energy between the states in the horizotally $HH$ and vertically $VV$ polarized photon emission cascades and $\Delta_t$ is generated randomly with an exponential distribution. 

On the other hand, the density matrix  $\rho_{pd}$ does not contain linear superpositions between differently polarized states but photons are still emitted correlated in polarization, 
\begin{equation}
\rho_{en}=\frac{1}{2}\left(\begin{matrix}
1 & 0 & 0 & 0 \\
0 & 0 & 0 & 0 \\
0 & 0 & 0 & 0 \\
0 & 0 & 0 & 1 
\end{matrix}
\right).
\end{equation}
The density matrix is therefore equivalent to a mixture of the states,
\begin{equation}
\begin{split}
&|\Psi\rangle=|HH\rangle\,,\\
&|\Psi\rangle=|VV\rangle\,,
\end{split}
\end{equation}
each one emitted with $1/2$ probability.

Finally, the noise density matrix $\rho_{noise}$,
\begin{equation}
\rho_{noise}=\frac{1}{4}\left(\begin{matrix}
1 & 0 & 0 & 0 \\
0 & 1 & 0 & 0 \\
0 & 0 & 1 & 0 \\
0 & 0 & 0 & 1 
\end{matrix}
\right)
\end{equation}
models the random emission of photons in any of the four possible pure basis states with equal probability $1/4$. These are,
\begin{equation}
\begin{split}
&|\Psi\rangle=|HH\rangle \\
&|\Psi\rangle=|HV\rangle \\
&|\Psi\rangle=|VH\rangle \\
&|\Psi\rangle=|VV\rangle 
\end{split}
\end{equation}

The process of generating a list of states compatible with the quantum dot emission density matrix consists in choosing the different paths of the next branching tree with the stated probabilities,
\begin{small}
\begin{equation}
|\Psi_i\rangle=
\begin{cases}
k \cdot p_s 	\begin{cases}
			p_d & |\Psi\rangle=|HH\rangle +  e^{-iS\Delta t/\hbar} | VV \rangle \\
			1 - p_d & \begin{cases} 
					  1/2 & |HH\rangle\\
					  1/2 & |VV\rangle
		    		      \end{cases} 
					  \\
			\end{cases}
			\\
1 -k \cdot p_s 	\begin{cases}
				1/4 & |HH\rangle \\
				1/4 & |HV\rangle \\
				1/4 & |VH\rangle \\
				1/4 & |VV\rangle 
				\end{cases}
\end{cases}
\end{equation}
\end{small}
The process is repeated each time a new pair of photons is emitted.

The output density matrix of a circuit obtained from stochastically sampling an imperfect source is calculated as the sum of all the individual output states,
\begin{equation}
\rho=\frac{1}{N}\sum_{i=1}^{N} |\Psi_i\rangle\langle \Psi_i |.
\end{equation}
Note that in the case of experiments with fluctuating emission parameters, it is needed to structure the output to avoid the infinite growth in the number of packets. That is,
if in each simulation the wavepackets are different, then after a large number of samples that number becomes also large. Fortunately, there are many ways to handle this situation.
If the detectors are counters, the particular packets, modeling the time of arrival of photons to those detectors, are irrelevant. 
On the other hand, detection may be performed or treated in a discrete manner. This discretization into intervals of detection reduces the number of possible outcomes in relation to time (or any other quantity)
to a finite manageable size.

A density matrix is represented in SOQCS by the C++ object \verb|dmatrix|. Pure states can be stored/added to \verb|dmatrix| like in a list but internally a matrix representation
is maintained. The matrix can be printed at any moment, updated with more states and printed again with improved statistics. Entries in the matrix (rows and columns indexes) are created dynamically to represent
only entries with non-zero values. Internally a dictionary is maintained to recognize if there is already a pre-existing entry to a particular coefficient that has to be updated or if a new entry (row and column) has to be created.

Note too that density matrices may also be needed in cases where the input to the circuit is a pure state. This may happen when post-selection conditions are multi-valued. 
Once again, we will expand upon this in section \ref{Detectors}.

\subsubsection{Entanglement swapping protocol}
In the entanglement swapping protocol \cite{Entanglement} two pairs of photons are emitted by a quantum dot in entangled Bell states at two different times. Each pair is emitted from a cascade where one of the photons is created from a bi-exciton XX and the other from an exciton X decay.

The exciton X photon of each pair is sent to a beamsplitter, each one into a different channel. The first exciton photon 
is delayed using a Mach-Zehnder interferometer. This can be modeled using a delay or, in this case, assuming for simplicity a delayed time as the emission time of the first exciton photon. If one photon is detected in each output channel of the beamsplitter then the remaining two photons become entangled. Ideally, this condition is only met when the two photons that arrive to the beamsplitter are in a $\Psi^-$ state. However, the quantum dot does not emit perfect Bell states and some fluctuations in the photons emission times also applies. Therefore, the resulting density matrix of the remaining bi-exciton XX photons will not correspond to a pure state of a perfectly entangled pair of photons. 

\begin{figure}[htb]
\centering
\includegraphics[width=0.5\textwidth]{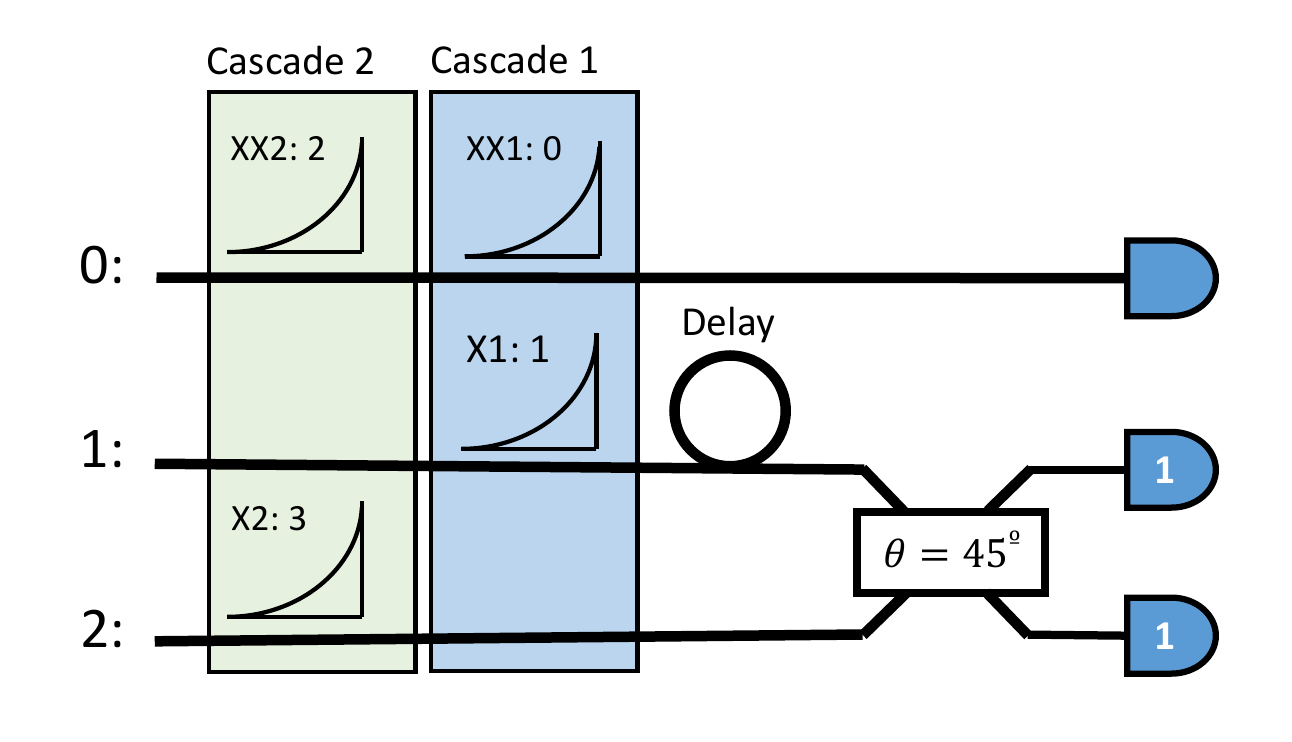}
\caption{Schematic of a circuit implementing the entanglement swapping protocol as discussed in ref. \cite{Entanglement}. XX are bi-exciton and X exciton photons emitted by a quantum dot modeled as an imperfect
source of photons. Detectors in channels one and two impose a post-selection condition of one photon detected in each channel.}.
\label{F7}
\end{figure}

This experiment can be modeled with the circuit in fig. \ref{F7} using the imperfect sources of photons. It is implemented in SOQCS using the following code: 
\begin{small}
\begin{verbatim}
for i in range(0,N,1):
    # Circuit creation
    example = 
    soqcs.qodev(4, 3, 2, 4, 1,ckind='E')
    
    # Imperfect QD emitter
    example.add_QD(0, 1,  
            t1=00.00, f1=10000.0, w1=1.0, 
            t2=46.71, f2=10000.0, w2=1.0, 
            PW=0, S=1.0, k=0.8, tss=1.0, 
            thv=1.0)
    example.add_QD(0, 2,  
            t1=16.00, f1=10000.0, w1=1.0, 
            t2=46.50, f2=10000.0, w2=1.0, 
            PW=0, S=1.0, k=0.8, tss=1.0, 
            thv=1.0)
            
    # Circuit
    example.beamsplitter(1,2,45.0,0.0);
    
    # Detectors
    example.detector(0)
    example.detector(1,1)
    example.detector(2,1)

    # Simulation
    inputst=example.input()
    circuit=example.circuit()
    outputst=sim.run_st(inputst,circuit)

    # Storage in a density matrix
    apd.add_state(outputst,example)
\end{verbatim}
\end{small}
where a circuit is created for exponentially shaped photon wavepackets and with sufficient number of degrees of freedom to perform the simulation.
The two cascades of photon are declared using imperfect sources of photons. In this simple simulation we assume the delay of the first exciton photon implicitly as part of the emission time. Here the protocol is simulated in the same way as is performed in  \cite{Entanglement} and depicted in fig. \ref{F7} where the two bi-exciton XX photons arrive at the same detector through channel zero but at different times that are labeled as 0 and 2. 

The simulation is repeated as many times as necessary with different initial states and the results are stored in the density matrix to reconstruct the statistics. 
The protocol is successful if two photons are measured at channels 1 and 2 independently of their detection time. On the other hand, the delay between photons in channel 0 is large enough to label them while disregarding small variations in their particular detection times among different simulations.

The density matrix below is obtained from repeated runs of the presented code. The result is consistent with analytical calculations using equivalent random and cross-dephasing noises and fine structure splitting values of the quantum dot while disregarding pure dephasing effects,

\begin{footnotesize}
\begin{verbatim}
| H(0)0, H(2)0 >  0.2251  0.0000  0.0000  0.0000 
| H(0)0, V(2)0 >  0.0000  0.2801 -0.0225  0.0000 
| H(2)0, V(0)0 >  0.0000 -0.0225  0.2703  0.0000 
| V(0)0, V(2)0 >  0.0000  0.0000  0.0000  0.2245 
\end{verbatim}
\end{footnotesize}

where the first column indicates the basis vectors corresponding to each row of the density matrix.

\subsection{Losses}
\subsubsection{Model}
Photon losses imply that the total probability of obtaining an outcome with the same number of photons as the input is smaller than one. If the simulator calculation only considers outputs with the same
number of photons as the input then output states are obtained normalized to values smaller than one. Therefore, a circuit with losses is described by a non-unitary matrix. An example can be provided by a lossy beamsplitter which is discussed below.

SOQCS implements a beamsplitter model of losses in which an extra virtual loss mode is added for each physical mode present in the circuit. The non-unitary matrix of the circuit is thus extended to include those extra degrees of freedom. Even if losses are defined locally (for each optical element) the loss model has to consider the entire circuit matrix. The reason is that we can not attribute a photon loss to any particular circuit element with certainty but all the possibilities have to be considered simultaneously and in linear superposition. Therefore, the matrix representing the entire circuit extended to include the virtual loss modes has to be unitary. If a photon is not transmitted then it is lost and has to appear in one of the virtual loss modes thus conserving probability. This way, we can use the already mentioned simulation methods for ideal circuits in which the number of photons at the output is the same as in the input. As a result, the implementation of the calculation of losses is independent of the chosen simulation method allowing us to achieve our goal of modularity.

It is always possible to construct a $2n \times 2n$ unitary matrix $U$ from a smaller $n \times n$ non-unitary matrix $M$ if the singular values of $M$ are smaller than one. The singular value decomposition can be
expressed as $M=RDV$ where $D$ is a diagonal matrix and $R$ and $V$ are unitary matrices. The $2n \times 2n$ matrix is built as,

\begin{equation}
\left(\begin{matrix}
M & R\sqrt{\left(I-D^2\right)}V  \\
  R\sqrt{\left(I-D^2\right)}V & -M \\
\end{matrix} \right).
\end{equation}

Finally, after a simulation, virtual loss channels can be traced out from the output state to obtain the circuit density matrix. This method can be applied independently of the circuit structure and parameters and it is used to obtain the amplitude of probability of $m$ photon outcomes given $n>m$ photon inputs. 

\subsubsection{Example: Lossy beasmplitter}
We can test the library for the particular case of a small thin dielectric film using the following equations \cite{Lossy}, 
\begin{align*} 
\hat{a}_{out}= t \hat{a}_{in}+ r  \hat{b}_{in} + \hat{F}_a  \\
\hat{b}_{out}= t \hat{b}_{in}+ r  \hat{a}_{in} + \hat{F}_b, 
\end{align*}
where we re-interpret the loss operators $\hat{F}_a$ and $\hat{F}_b$ as virtual channel destruction operators (or at least their normalized counterparts $\hat{F}_a=\sqrt{1-|r|^2-|t|^2}\hat{f}_a$). $\hat{F}_a$ and $\hat{F}_b$ do not commute and their self commutators are not one,
\begin{align*} 
[\hat{F}_a,\hat{F}_a^\dagger]&=[\hat{F}_b,\hat{F}_b^\dagger]=1-|t|^2-|r|^2 \\
[\hat{F}_a,\hat{F}_b^\dagger]&=-tr^*+rt^* .
\end{align*}
This thin dielectric act as an imperfect beamsplitter with losses. The dielectric is modeled as the non-unitary gate,
\begin{equation}
\left(\begin{matrix}
t & r \\
r & t \\
\end{matrix} \right),
\end{equation}
that is extended following the previously established  prescription,
\begin{equation}
\left(\begin{matrix}
t & r  & \lambda_+ & \lambda_-\\
r & t & \lambda_- & \lambda_+\\
\lambda_+ & \lambda_- & -t & -r \\
\lambda_- &  \lambda_+ & -r & -t \\
\end{matrix} \right),
\end{equation}
where $\lambda_\pm=\frac{1}{2}(\sqrt{1-|r+t|^2}\pm\sqrt{1-|r-t|^2})$. 

For this particular case, the dielectric matrix is normal therefore $R=V^\dagger$. $D$ is the eigenvalue matrix and $R$ contains the eigenmodes of the circuit. Hence, in this case we can interpret $\sqrt{1-D^2}$ as the coupling coefficients between each of those eigenmodes and a single loss mode.

\begin{figure}[h]
\centering
\includegraphics[width=0.45\textwidth]{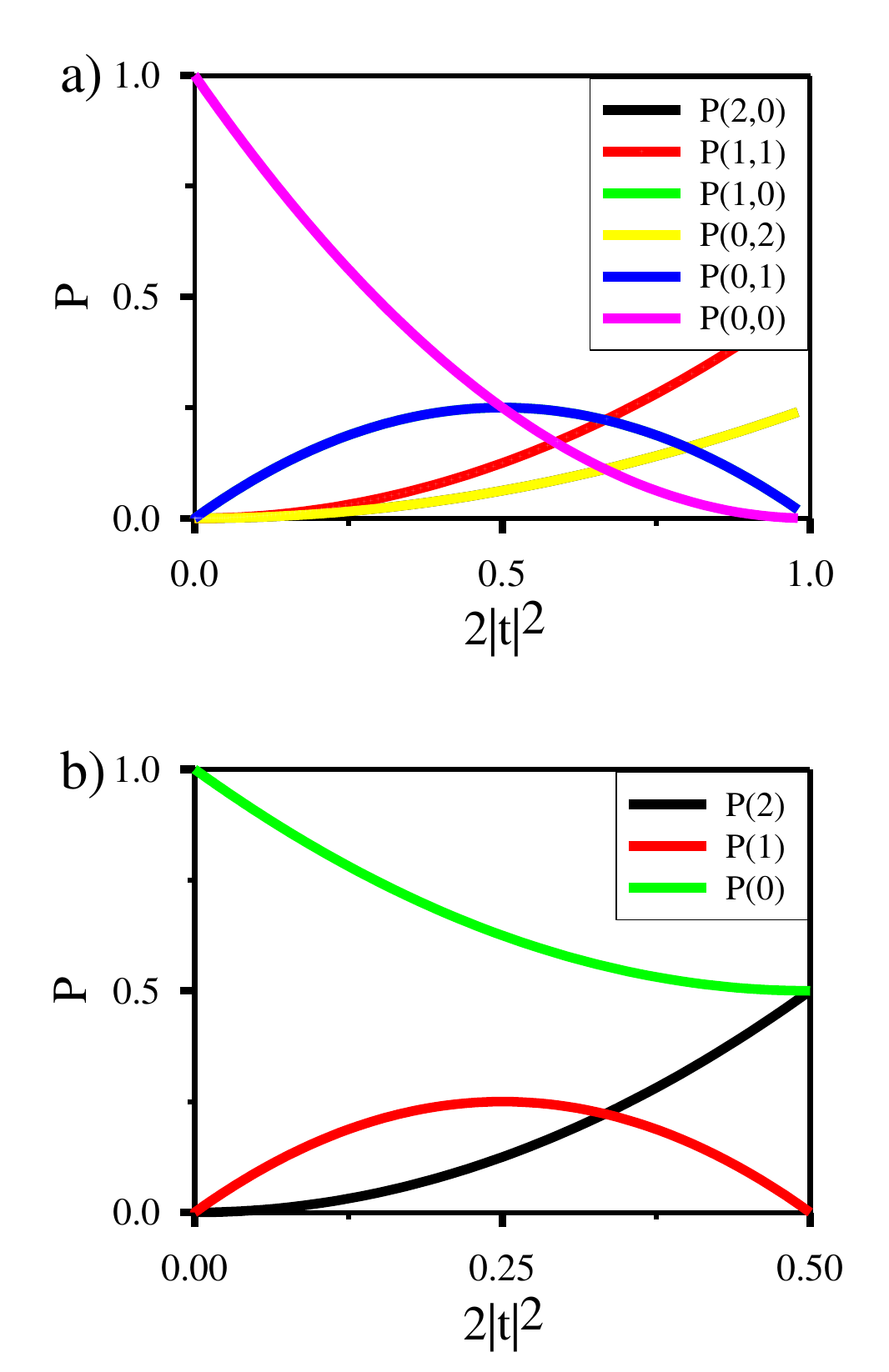}
\caption{Outcome probability of a lossy beamsplitter a) in case that two identical photons are set-up in the same input channel $|2,0\rangle$ and $t=ir$. b) in case that two identical photons are set up, one in each input channel, $|1,1\rangle$ and $t=r$. Both calculations carried out in SOQCS reproduce the results for the same device presented in ref. \cite{Lossy}. }
\label{F8}
\end{figure}

The analytical result presented in ref. \cite{Lossy} for different transmission amplitudes can be reproduced numerically in SOQCS using this matrix as shown in fig. \ref{F8}.

\section{Detectors}
\label{Detectors}
SOQCS handles detectors as virtual circuit elements that can be defined in the same manner as the linear optical elements. However, the process of detection is a non-unitary irreversible operation. Their virtual definitions are used to specify the parameters needed to configure the post-selection conditions of the circuit and the different models of imperfections in the detection process.

\subsection{Post-selection}
In general, the probability of measurement of a determined outcome is,
\begin{equation}
P(i)=Tr[\Pi_i \hat{\rho}],
\end{equation}
where $\hat{\rho}$ is the density operator and $\Pi_i=|\Phi_i \rangle\langle \Phi_i| $ is the projector operator. Post-selection is a partial measurement where only some of the degrees of freedom that
describe a state are measured. If a certain outcome is fulfilled then the output obtained in the remaining modes is accepted (otherwise it is rejected). Thus, the result of post-selection is a reduced 
density matrix for the remaining degrees of freedom,
\begin{equation}
\tilde \rho_i=Tr_{\vec \omega}[(I\otimes\Pi_{\omega_i}) \hat{\rho} (I\otimes\Pi_{\omega_i})].
\end{equation}

If the output of a circuit is calculated repeatedly in presence of noise or with different inputs (ex. non ideal source) then we will obtain a different output each time. In this case, the original unreduced density matrix can be re-interpreted as the average of density matrices of each of the output pure states  $\hat{\rho}=\frac{1}{N}\sum_{j=1}^{N}\rho_j$. As a consequence, the reduced density matrix can be rewritten as the average of each pure state matrix reduced by post-selection,
\begin{equation}
\tilde \rho=\sum_{j=1}^{N} Tr_{\vec \omega}[(I\otimes\Pi_{\omega_i}) \rho_j (I\otimes\Pi_{\omega_i})].
\end{equation}

This means that we can calculate the overall reduced density matrix as the sum of already post-selected individual state projectors,
\begin{equation}
\tilde \rho=\frac{1}{N}\sum_{j=1}^{N} | \tilde\Psi_j\rangle \langle \tilde \Psi_j |.
\end{equation}

Therefore to perform post-selection in SOQCS,
\begin{itemize}
\item First, the output state of an optical circuit is obtained 
\item Second, post-selection is performed applying a projector to the output state ( defined for a subset of degrees of freedom of that state). 
\item Third, the resulting reduced state is added to a density matrix in the same manner as it is done in circuits with no post-selection.
\end{itemize}

We can interpret the post-selection operation as a function that relates a state with another state defined in a reduced set of degrees of freedom. For example, if we perform a post-selection operation over the state, 
\begin{equation}
\begin{split}
|\Psi\rangle_{0123}=&\frac{1}{2}( |\Phi^+\rangle_{03}|	\Phi^+\rangle_{12} \\
&+ |\Phi^-\rangle_{03}| \Phi^-\rangle_{12} \\
&+ |\Psi^+\rangle_{03}|\Psi^+\rangle_{12} \\
&+ |\Psi^-\rangle_{03}|	\Psi^-\rangle_{12}  )\,,
\end{split}
\end{equation}
projecting the 2nd and 3rd degrees of freedom (usually channels) to  the projector $|\Psi^-\rangle_{12}$ the resulting reduced state is $|\tilde\Psi\rangle=|\Psi^-\rangle_{03}$. 

However, post-selection is not always defined for a particular projector. It is usual to demand certain conditions (for example occupation values) to be fulfilled in certain channels independently of their polarization, frequency or other characteristics. In this case more than one projector will fulfill the output condition. For example, in the entanglement swapping protocol explained above the post-selection condition is the detection of one photon in both channels one and two  independently of photon polarization. Therefore, the projectors considered can be $|HH\rangle,|HV\rangle,|VH\rangle,|VV\rangle$ where  photons may be "vertically" $V$ or "horizontally" $H$ polarized or, alternatively, $|\phi^+\rangle,|\phi^-\rangle,|\psi^+\rangle,|\psi^-\rangle$. The reduced density matrix is calculated as the sum of reduced density matrices obtained from applying each projector,
\begin{equation}
\tilde \rho=Tr_{\vec \omega}[\sum_{i=1}^{N}  ((I\otimes\Pi_{\omega_i}) \hat{\rho} (I\otimes\Pi_{\omega_i}))].
\end{equation}
SOQCS determines automatically the projectors needed, given the output condition, and performs the calculation of the density matrix automatically.

\subsection{Detector abstraction}
Measurement may depend on the kind of detector used. For example, if the detector is only a counter of photons to be read at the end of the experiment then that detector is blind
to the particular characteristics of the photon wavepackets (time, frequency, etc). Therefore, the degrees of freedom of the resulting density matrix have to be reduced to obtain a new state description without the packet degree of  freedom. This does not mean that the packet degree of freedom is unimportant in the circuit outcome. For example, in fig. \ref{F6} we can see the probability of having five photons in two different channels of a circuit where detectors are configured as counters. 
The counting in different channels is different depending on whether they are synchronized due to HOM. 
A relabeling of the output to remove undesired degrees of freedom is always possible for the diagonal elements of the density matrix while off diagonal elements are more involved and depend on the particular detector definition.

In general, there are three kinds of detectors defined in SOQCS,
\begin{itemize}
\item {\bf Counters}. All photons in a mode are summed independently of their packet degrees of freedom.
\item {\bf Timed detection}. The packet index is dropped in favor of a time label. 
\item {\bf Full detection}. We do not perform any further process to the output and the result is expressed in terms of the orthonormalized packets.
\end{itemize}

Configuring a quantum device to use the proper kind  of detector will provide automatically the desired probability distribution (and in some cases also the density matrix).

\subsection{Imperfect physical detectors}
Detectors are physical devices with their own imperfections. Detector imperfections affect the outcome probabilities of a circuit. For example, the {\bf detector efficiency} is related to losses within the detector. In SOQCS, this can be calculated automatically by adding lossy medium to the circuit when the detector is defined.

With the exception of losses the rest of the detection imperfections can be calculated with classical error models. These errors alter the distribution of probability of the circuit outcomes. (Distribution probabilities are the diagonal coefficients of the density matrix. Errors are undefined for the off diagonal coefficients because they have no classical meaning). In SOQCS there is a second C++ class that handles distributions of probabilities in a equivalent way to the diagonal elements of the density matrix. The probability bins class \verb|p_bin| is more efficient and faster if we are uninterested in the coherences for a particular problem. It is in this class where physical detector effects are implemented.

The classical error models implemented in SOQCS are,
\begin{itemize}
\item {\bf Dead time}. After a photon measurement the detector is off for a certain time. We model this effect by rejecting a detector measurement randomly with a certain probability. We define this probability as the fraction of time that a detector is off with respect to the simulation time. This model provides a very simple approximation. We are assuming that the main cause of dead time is noise with no correlation between detectors. This is normally true if the simulation time is large enough in comparison with the dead time. 
\item {\bf Dark counts}. A detector may observe photons emitted by itself. The dark count rate is the average rate of registered counts without any incident light. These false detections are mainly of thermal origin. Therefore, photons are added to the resulting outcome with a probability determined by a Poissonian distribution, 
\begin{equation}
P(n)=\frac{\lambda^n e^{-\lambda}}{n!}
\end{equation}
where $n$ is the number of photons (dark counts) to be added and $\lambda$ is the rate (average value) of dark counts in the simulation time. 
\item {\bf Noise}. Finally, the output probability may be altered with a random Gaussian noise to account for a wide range of sources of noise present in experimental setups.
\end{itemize}

The overall detection procedure computed in SOQCS is implemented as,
\begin{enumerate}
\item Calculate the probability distribution.
\item Compute dark counts.
\item Compute dead time.
\item Calculate post-selection.
\item Remove degrees of freedom depending on detector definitions.
\item Add noise.
\end{enumerate}

Note that the computation of losses adds some auxiliary extra loss modes. These modes can be eliminated from the probability distribution in step 4th performing a post-selection procedure
for all the possible outcomes in those channels.

\section{Conclusions}
\label{Conclusions}
In this paper we report the implementation of the optical circuit simulator SOQCS that uses an approach akin to stochastic quantum trajectories.
The simulator has been built in a modular way and allows for multiple simulation methods.
These methods provide as results output states, probability distributions and density matrices depending on user's configuration. Post-selection conditions and measurement configurations are implemented using detectors as virtual circuit elements. Furthermore, partial distinguishability, delays, losses, circuit and detector imperfections are parametrized and calculated automatically. Simulations can be carried out using SOQCS as a C++ library or by means of its Python port that also contains a graphical circuit plotter. In conclusion, SOQCS is a framerwork for the simulation of quantum optical circuits that has been built with flexibility in mind and with a modular code that can be further expanded to include additional simulation methods or new imperfection models.

\section*{Acknowledgements}
This work has received funding from the Enterprise Ireland’s DTIF programme of the Department of Business, Enterprise and Innovation, project QCoIr Quantum Computing in Ireland: A Software Platform for Multiple Qubit Technologies No. DT 2019 0090B. We acknowledge discussions and support from Tyndall National Institute and Rockley Photonics Ltd. We also thank Paul Watts for useful comments on the manuscript.

\bibliographystyle{elsarticle-num} 
\bibliography{A2Refs}

\appendix
\onecolumn
\newpage
\section{List of elements of a quantum device}
List of optical elements provided by the class \verb|qodev|. \verb|qodev| provides and abstract representation of a quantum optical circuit, its initialization state and helps handling the packet definitions that describe
the photon wavepackets.
\newline
\newline
{\bf Basic elements}
\begin{itemize}
\item \verb|void random_circuit()|. Random circuit.
\item \verb|int NSX(int i_ch1, int i_ch2, int i_ch3)|. NSX gate.
\item \verb|int beamsplitter(int i_ch1, int i_ch2, double theta, double phi)|. Beamsplitter.
\item \verb|int dielectric(int i_ch1, int i_ch2, cmplx t, cmplx r)|. Thin dielectric film.
\item \verb|int MMI2(int i_ch1, int ch2)|. 2x2 MMI.
\item \verb|int rewire(int i_ch1,int i_ch2)|. Swaps two channels.
\item \verb|int phase_shifter(int i_ch, double phi)|. Phase shifter.
\item \verb|int loss(int i_ch, double l)|. Lossy medium
\item \verb|int delay(int ch, double dt)|. Delay.
\end{itemize}

\phantom{-} 
\newline
{\bf Polarization elements}
\begin{itemize}    
\item \verb|int rotator(int i_ch, double d_theta, double d_phi)|. Polarizarion rotator.
\item \verb|int polbeamsplitter(int i_ch1, int i_ch2, int P)|. Polarizing beamsplitter.
\item \verb|int half(int i_ch, double alpha)|. Half-waveplate.
\item \verb|int quarter(int i_ch, double alpha)|. Quarter-waveplate.
\end{itemize}

\phantom{-} 
\newline
{\bf Detection elements}
\begin{itemize}
\item \verb|int ignore(int i_ch)|. Flags a channel to be ignored
\item \verb|int detector(int i_ch)|. Adds a detector
\item \verb|int detector(int i_ch, int cond)|. Adds a conditional detection.
\item \verb|int detector(int i_ch, int cond, double eff, double blnk, double gamma)|. Adds a general physical detector
\item \verb|void noise(double stdev2)|. Adds noise to the output
\end{itemize}

\newpage

\section{Example: Controlled phase flip gate}
\begin{figure}[h]
\centering
\includegraphics[width=1.0\textwidth]{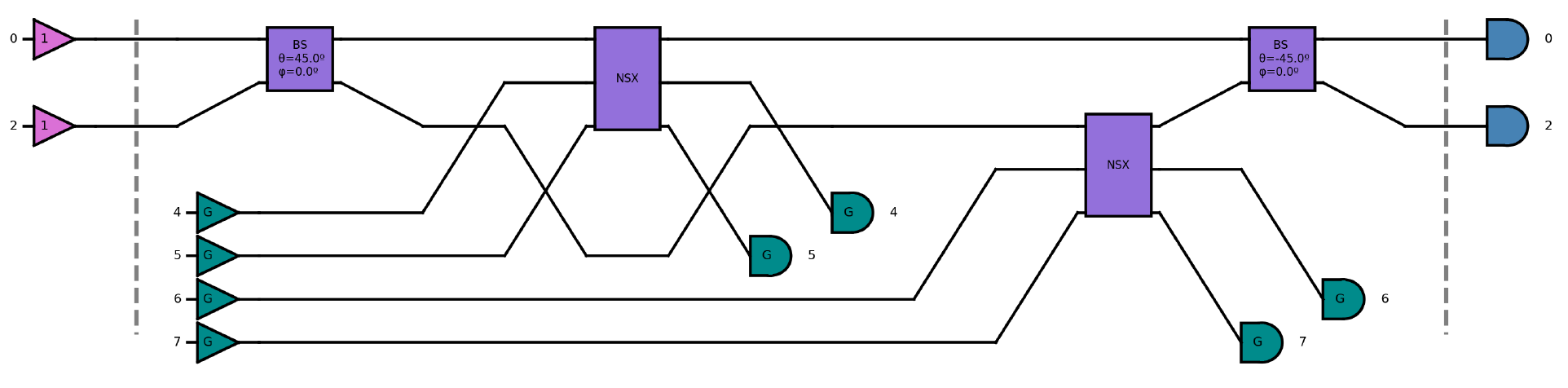}
\caption{The CZ circuit \cite{KLM}. Plot is generated automatically by SOQCS. The modes $1$ and $3$ are not acted upon by the circuit and hence are omitted from the plot to avoid a clutter.}
\label{F9}
\end{figure}

We use SOQCS to simulate a controlled phase flip gate CZ (see fig. \ref{F9}) as described in \cite{KLM}. We define and additional $NS_{-1}$ gate to build the CZ circuit. Note that the $NS_{-1}$ can be found
as part of SOQCS gate catalog. However, we prefer to define here the gate as an additional circuit instead of using the catalog version to show how circuits can be used to define new gates. This way custom libraries
of gates can be built. Channel 0 is left open to be connected in the larger CZ circuit while ancilla channels 1 and 2 are set-up to hold one and zero photons respectively. These same channels are post-selected to values 1 and 0 as an accepting condition.

\begin{verbatim}
# Create circuit
NSx = soqcs.qodev(4, 3)
NSx.open_channel(0)
NSx.add_photons(1,1)
NSx.add_photons(0,2)

# Build gate
NSx.phase_shifter(0,180.0)
NSx.beamsplitter(1,2,22.5,0.0)
NSx.beamsplitter(0,1,65.5302,0.0)
NSx.beamsplitter(1,2,-22.5,0.0)

# Configure post-selection detectors
NSx.detector(1,1)
NSx.detector(2,0)
\end{verbatim}

Next is defined the CZ circuit. The output is encoded as a qubit. Channels 0 and 1 define the first qubit while channels 2 and 3 the second. Qubits are path encoded, therefore, a photonic configuration "01" is equivalent to a qubit logical "0" and a configuration "10" to a logical "1". The circuit is initialized with the state $|1,0,1,0\rangle_{0,1,2,3}$ equivalent to a qubit encoded input $|1,1\rangle$.  Post-selection has already been defined as part of the $NS_{-1}$ gate.
\begin{verbatim}
# Define qubit map
qmap=[[0, 2],
      [1, 3]]

# Create circuit
csign10 = soqcs.qodev(4, 8)
#Initialize qubit channels
csign10.add_photons(1,0)
csign10.add_photons(1,2)
csign10.separator()

# Build cicuit
csign10.beamsplitter(0,2,45.0,0.0)
chlist = [ 0, 4, 5 ]
csign10.add_gate(chlist,NSx,text="NSX")
chlist = [ 2, 6, 7 ]
csign10.add_gate(chlist,NSx,text="NSX")
csign10.beamsplitter(0,2,-45.0,0.0)
csign10.separator()

# Configure detectors
csign10.detector(0)
csign10.detector(1)
csign10.detector(2)
csign10.detector(3)
\end{verbatim}

In this particular case, we are interested in a phase flip, and hence we instruct SOQCS to provide a state as output. Other options like probability distributions or density matrices
are available using different instructions.
\begin{verbatim}
# Run optical circuit simulation
raw_output=sim.run_st(csign10.input(),csign10.circuit())
# Apply the post-selection condition
output=csign10.apply_condition(raw_output)
# Encode the output in qubit form.
encoded=output.encode(qmap,csign10.circuit())
# Print the result
encoded.prnt_state(column=1)
\end{verbatim}

The intended output is shown next. Note that the norm of the qubit encoded state is informing us of the success probability of the circuit $p=1/16$.
\begin{verbatim}
| 1, 1 >: -0.25000000 + 0.00000000 j
\end{verbatim}

We can also use SOQCS in a different manner to set-up the input state $\frac{1}{2}(|0,0\rangle+|0,1\rangle+|1,0\rangle+|1,1\rangle)$ and perform the simulation,
\begin{verbatim}
# Create initial state
qubit= soqcs.state(2)
qubit.add_ket(0.5,[0,0])
qubit.add_ket(0.5,[0,1])
qubit.add_ket(0.5,[1,0])
qubit.add_ket(0.5,[1,1])

# Run the simulation for that state
raw_output=sim.run_st(qubit.decode(qmap,[1,0,1,0],csign10.circuit()),csign10.circuit())
# Apply the post-selection condition
output=csign10.apply_condition(raw_output)
# Encode the output in qubit form.
encoded=output.encode(qmap,csign10.circuit())
# Normalize the output
encoded.normalize()
# Print the result
encoded.prnt_state(column=1)
\end{verbatim}
The resulting output is shown next. Note that in this case we have instructed SOQCS to provide a normalized output,
\begin{verbatim}
 | 0, 0 >:  0.50000000 + 0.00000000 j
 | 0, 1 >:  0.50000000 + 0.00000000 j
 | 1, 0 >:  0.50000000 + 0.00000000 j
 | 1, 1 >: -0.50000000 + 0.00000000 j
\end{verbatim}

\newpage

\section{Example: CNOT gate}
\begin{figure}[h]
\centering
\includegraphics[width=1.0\textwidth]{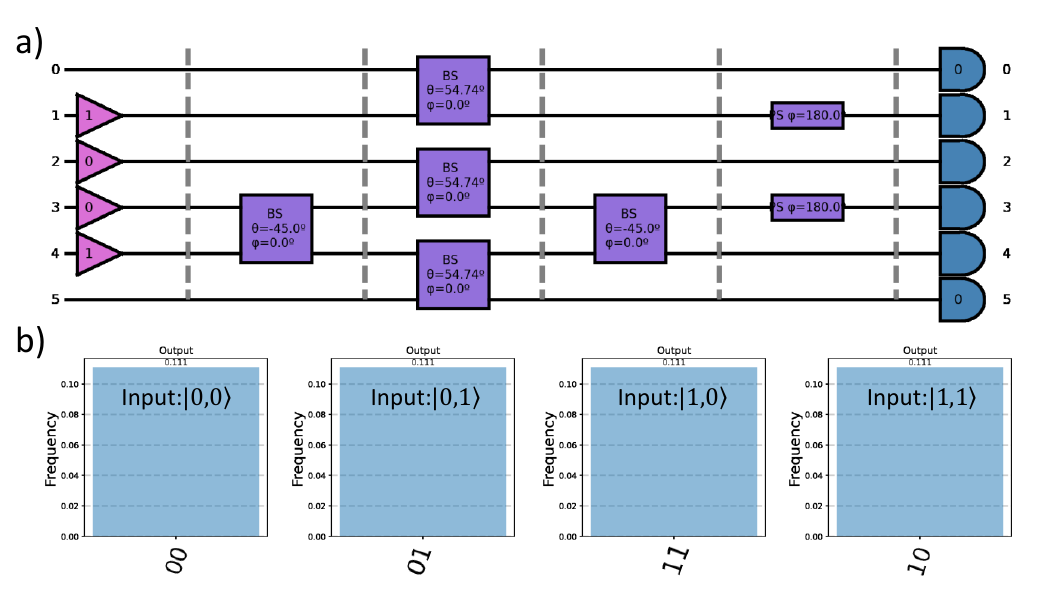}
\caption{ a) CNOT circuit as described in ref. \cite{CNOT}. In this example, the two-qubit input state is $|1,0\rangle$, encoded into the modes $1$ to $4$. The result is postselected on the zero photon states in the ancillary modes $0$ and $5$.
b) Output of the circuit in qubit representation as provided by SOQCS for the inputs $|0,0\rangle$, $|0,1\rangle$, $|1,0\rangle$ and $|1,1\rangle$ respectively. The success probability of the circuit is $1/9$ as indicated by the height of the plots.}
\label{F10}
\end{figure}

We use SOQCS to simulate a circuit of a probabilistic implementation of the CNOT gate as described in \cite{CNOT}, see also fig. \ref{F10}a. Channels 1 and 2 define the first qubit while channels 3 and 4 the second. Qubits are path encoded in the same way as in the CZ example. However, in this case the encoding is used also to directly define the input $|1,0\rangle$. The code to build the CNOT circuit is provided below. Ancilla channels are initialized to zero photons.

\begin{verbatim}
# Define qubit map
qmap=[[1, 3],
      [2, 4]]
      
# Create circuit      
cnot10=soqcs.qodev(2,6); 
# Initialize input
cnot10.qubits([1,0],qmap) 

# Build circuit
cnot10.beamsplitter(3,4, -45.0,0.0)
cnot10.beamsplitter(0,1,180*acos(1.0/sqrt(3.0))/pi,0.0)
cnot10.beamsplitter(2,3,180*acos(1.0/sqrt(3.0))/pi,0.0)
cnot10.beamsplitter(4,5,180*acos(1.0/sqrt(3.0))/pi,0.0)
cnot10.beamsplitter(3,4, -45.0,0.0)
cnot10.phase_shifter(1, 180)
cnot10.phase_shifter(3, 180)
cnot10.detector(0,0)
cnot10.detector(1)
cnot10.detector(2)
cnot10.detector(3)
cnot10.detector(4)
cnot10.detector(5,0)
\end{verbatim}
Here, SOQCS is instructed to obtain the output distribution automatically from the output state. The result is plotted graphically and presented in fig. \ref{F10}b.
\begin{verbatim}
outcome=sim.run(cnot10)
encoded=outcome.translate(qmap, cnot10)
encoded.show(sizex=5,dpi=70)
\end{verbatim}

\end{document}